\documentclass[twocolumn,american,showpacs]{revtex4}
\usepackage{graphicx}
\usepackage{amsmath}
\usepackage{amssymb}
\usepackage{hyperref}

\newcommand{\eg}{{\it eg~}}
\newcommand{\et}{{\it et al~}}

\newcommand{\s}{{Section~}}
\newcommand{\e}{{Eq.~}}
\newcommand{\es}{{Eqs.~}}
\newcommand{\A}{{Appendix~}}

\def\mathbi#1{\textbf{\em #1}}

\begin{document}

\title{Modelling of nematic liquid crystal display devices}

\author{Spencer, T.J. and Care, C.M. }

\affiliation{Materials $\&$ Engineering Research Institute,
Sheffield Hallam University, Howard Street, Sheffield, S1~1WB,
United Kingdom}

\begin{abstract}

A lattice Boltzmann  scheme is presented which recovers the dynamics
of nematic and chiral liquid crystals; the method essentially gives
solutions to the Qian-Sheng \cite{Qian:1998.7475} equations for the
evolution of the velocity and tensor order-parameter fields. The
resulting algorithm is able to include five independent Leslie
viscosities, a Landau-deGennes free energy which introduces three or
more elastic constants, a temperature dependent order parameter,
surface anchoring and viscosity coefficients, flexo-electric and
order electricity and chirality. When combined with a solver for the
Maxwell equations associated with the electric field, the algorithm
is able to provide a full \lq device solver' for a liquid crystal
display. Coupled lattice Boltzmann schemes are used to capture  the
evolution of the fast momentum and slow director motions in a
computationally efficient way. The method is shown to give results
in close agreement with analytical results for a number of
validating examples.  The use of the method is illustrated through
the simulation of the motion of defects in a zenithal bistable
liquid crystal device.

\end{abstract}


\maketitle

\section{Introduction}
\label{intro}

Many of the next generation of liquid crystal (LC) display devices
use structured or patterned
surfaces as an essential element of their design and function \cite{patent, Kim:2001.3055,
Schift:2002.423}. The correct operation of these devices depends
upon the formation and annihilation of defects in the orientational
field of the nematic; the defects are usually intimately coupled to
a  surface.  There is also current interest in the behaviour of LC's
with embedded colloidal particles (\eg
\cite{Stark:2001.387,Loudet:2004.11336}); the behaviour of these
materials is frequently dependent upon the interaction of the
defects associated with the colloidal particles.

Experimentally it is difficult to obtain information about the
spatial and temporal behaviour of the nematic order. Optical methods
such as those of~\cite{Jewell:2002.19, Jewell:2003.3156} can, for
example, give information about the director profiles on a
relatively coarse time and space scale. However, in order to fully
understand such systems it is necessary to be able to model the
statics and dynamics of a nematic in the presence of complex
boundaries and defects.  The problem is compounded by the numerous
materials parameters needed to fully describe the properties of the
LC and its interaction with any bounding surfaces; predictive
modelling often requires a fairly complete description of the
materials and this therefore necessitates the use of numerical
methods to solve the associated equations.  In this paper we present
the details of one such numerical method and illustrate its use with
a number of examples.

LC's are complex fluids formed from anisometric molecules. These
fluids can exhibit a range of mesophases with varying degrees of
orientational and positional order of the molecules; in the nematic
phase there is long range orientational order but no positional
order.  The orientational ordering is described at mesoscopic length
scales by an order tensor, the Q-tensor (see \s \ref{model}), whose
principal eigenvalue is related to the order parameter  and whose
principal eigenvector defines the macroscopic director field
\eg\cite{deGennes:1969.454,deGennes:1971.193,deGennes:1993}. In many
systems it is possible to assume that the order parameter is
constant and the dynamics of the momentum and director are then
described by the well established Ericksen-Leslie-Parodi (ELP)
equations (\eg \cite{deGennes:1993}).

However, near to bounding walls and close to defects, the assumption
of constant order parameter breaks down and the material may also
exhibit biaxiality.  There are significant spatial gradients in the
order tensor in such regions and the gradients have observable
macroscopic consequences. For example, Q-tensor gradients lead to
the flexo- and order-electric polarisation which is used to control
the switching behaviour of some display devices by an applied
electric field. Similarly, the dynamics of defects can only be
correctly described within a theoretical framework which allows for
variation in the order parameter.

In such systems it is necessary to go beyond the ELP theory and
adopt a model which describes the dynamics of the full Q-tensor.
There are a number of derivations of nematodynamics with variable
order parameter (\eg
\cite{Hess:1975.728,Olmsted:1990.4578,Beris:1994,Qian:1998.7475}).
Work by Sonnet \et \cite{Sonnet:2004.51} provides the basis upon
which the variety of schemes with a variable order parameter may be
compared; it should be noted that in the limit that the order
parameter becomes independent of time and position all the schemes
must recover the ELP theory. In this work we adopt the Qian-Sheng
\cite{Qian:1998.7475} formalism because it is straightforward to
obtain the required material parameters from those of the equivalent
ELP description. Exact solutions to either the ELP or Q-tensor
theories are limited to a relatively small number of simplified
cases and numerical methods are necessary for the more complex
systems which form the focus of this work.

A number of different approaches have been taken to finding
numerical  solutions to the equations for variable order parameter
nemato-dynamics. Svensek~\cite{Svensek:2002.021712} and
Fukuda~\cite{Fukuda:2004.S1957}  use conventional methods to solve
the associated partial differential equations.  However, a number of
workers have adapted the lattice Boltzmann (LB) method (\eg
\cite{Care:2005.2665,Care:2000.L665,Denniston:2000.481,Care:2003.061703,Denniston:2004.1745}).
Care \et~\cite{Care:2000.L665} developed the first methods in which
LB was used to solve the ELP equations in a 2-D plane, and
later~\cite{Care:2003.061703} enhanced the method to yield a two
dimensional solution to the steady state Qian-Sheng equations;
concurrently, Denniston \et~ (\eg
\cite{Denniston:2000.481,Denniston:2001.389,Denniston:2004.1745})
developed LB tensor methods for nematic liquid crystals based on the
Beris-Edwards~\cite{Beris:1994} scheme.

The LB method may be regarded simply as an alternative method of
solving a target set of macroscopic differential equations. However,
it is advantageous to regard it as a mesoscale method which allows
additional physics to be included within the modelling; this is
illustrated by the extension of the method to model  the interface
between an isotropic and nematic fluid \cite{Care:2003.061703}, a
problem of direct relevance to modelling liquid crystal colloids. LB
has the additional stability advantages of being able to incorporate
complex boundary conditions more easily than conventional solvers
and being straightforward to parallelise.

In this paper we present an LB scheme which recovers the Qian-Sheng
equations for nemato-dynamics. The approach  modifies the scheme
presented in \cite{Care:2003.061703} by utilising a simple LBGK
scheme (\eg \cite{Qian:1992.479}) for the collision term and
introducing all the anisotropic behaviour through forcing terms.
This moves away from the goal which was implicit in
\cite{Care:2003.061703} of remaining as close as possible to the
physical basis of nematodynamics by using an anisotropic collision
operator. However, the overhead in numerical complexity made the
scheme difficult to generalise to three dimensions, a restriction
which does not apply to the method presented in this paper.

The resulting algorithm includes five independent Leslie
viscosities,  a Landau-deGennes free energy which introduces three
or more elastic constants, a temperature dependent order parameter,
surface anchoring and viscosity coefficients, flexo-electric and
order electricity and chirality. The precise properties of the
system, such as the number of elastic constants, is modified by the
inclusion or exclusion of terms in the free energy. When combined
with an appropriate solver for the electric field, the algorithm is
able to provide a full \lq device solver' for a liquid crystal
display. The method employs two lattice Boltzmann schemes, one for
the evolution of the momentum and one for the evolution of the
Q-tensor. This is necessary because of the large differences in time
scale for the evolution of  velocity and director fields in a
typical display or experimental arrangement.
The paper is organised as follows. In \s\ref{model} we outline the
Qian-Sheng equations for tensor nemato-dynamics. In \s
\ref{algorithm} we present an LB method to recover these equations.
In \s\ref{results}, results are presented for the validation of the
method against analytical equations and  the application of the
method to the modelling of a Zenithally Bistable Device (ZBD) device
is reported. \s\ref{conclusions} concludes by highlighting the
benefits of the method described in the paper and discusses the
implications for future work.  A Chapman-Enskog analysis (\eg
\cite{Chapman:1995}) to justify the LB scheme is given in \A
\ref{algorithm_analysis} and \A \ref{parameters} summarises useful
relationships between the vector, tensor and LB parameters of the
method and lists the material constants used in the simulations
detailed in \s\ref{results}.


\section{The Qian-Sheng formalism}
\label{model}

In this section we summarise the Qian-Sheng formalism
\cite{Qian:1998.7475} for the flow of a nematic liquid crystal with
a variable scalar order parameter. The tensor summation convention
is assumed over repeated greek indices which represent three
orthogonal Cartesian coordinates; no summation convention is assumed
for roman indices which are used to indicate lattice directions of
the LB algorithm. $\delta_{\alpha \beta}$ and $ \varepsilon_{\alpha
\beta \gamma}$ are the Kronecker delta and Levi-Civita symbols
respectively and a superposed dot (~$^{\dot{}}$~) denotes the
material time derivative: $\partial _{t}+u_{\alpha }\partial
_{\alpha}$.

The symmetric, and traceless, macroscopic order tensor,
$\mathbi{Q}$, is defined to be
\begin{equation}\label{eq_q_tensor}
    Q_{\alpha \beta}=\frac{S}{2}\left( 3\hat{n}_{\alpha }\hat{n}_{\beta}-\delta _{\alpha \beta } \right) + \frac{P_{_{B}}}{2}\left( \hat{l}_{\alpha }\hat{l}_{\beta}-\hat{m}_{\alpha }\hat{m}_{\beta } \right)
\end{equation}
where $S$ and $P_{_{B}}$ are the uniaxial and biaxial order
parameters with $\hat{\mathbi{n}}, \hat{\mathbi{l}}$ and
$\hat{\mathbi{m}}$ being orthogonal unit vectors associated with the
principle axes of $\mathbi{Q}$. In the uniaxial approximation
$P_{_{B}}=0$ and in the ELP approximation the scalar order parameter
$S\rightarrow S_{0}$, a constant.  The director, $\hat{\mathbi{n}}
=(\sin \theta \cos \phi ,\sin \theta \sin \phi ,\cos \theta )$, is
the eigenvector corresponding to the largest eigenvalue of
$\mathbi{Q}$.

Following ~\cite{Qian:1998.7475}, the momentum and order evolution
equations for incompressible ($\partial _{\alpha }u_{\alpha }=0$)
nemato-dynamics are written as
\begin{equation}\label{eq_qs_momentum}
    \rho \dot{u} _{\beta } = \partial _{\alpha }\left( -P\delta _{\alpha \beta }+\sigma _{\alpha \beta }^{v}+\sigma _{\alpha \beta }^{d}+\sigma _{\alpha \beta }^{EM} \right)
\end{equation}
\begin{equation}\label{eq_qs_order}
    J\ddot{Q}_{\alpha \beta }=h_{\alpha \beta }+h_{\alpha \beta }^{v}-\lambda \delta _{\alpha \beta }-\varepsilon _{\alpha \beta \gamma }\lambda _{\gamma }
\end{equation}
Here the local variables are $\rho $ the liquid crystal density,
$\mathbi{u}$ the fluid velocity, $P$ the pressure, $J$ the moment of
inertia.  $\lambda $ and $\lambda _{\gamma }$ are Lagrange
multipliers chosen to ensure that $\mathbi{Q}$ remains symmetric and
traceless. $\sigma _{\alpha \beta }^{d}$ and $h_{\alpha \beta }$ are
the distortion stress tensor and molecular field defined by the
Landau-deGennes free energy, $F$, for the system through the
expressions
\begin{equation}\label{eq_def_dist_stress}
    \sigma _{\alpha \beta }^{d} = -\frac{\partial F_{Bulk}}{\partial (\partial _{\alpha }Q_{\mu \nu })}\partial _{\beta }Q_{\mu \nu}
\end{equation}
\begin{equation}\label{eq_def_mol_feild}
    h_{\alpha \beta }=-\frac{\partial F_{Bulk}}{\partial Q_{\alpha \beta }}+\partial _{\gamma }\frac{\partial F_{Bulk}}{\partial (\partial _{\gamma }Q_{\alpha \beta })}
\end{equation}
$\sigma _{\alpha \beta }^{v}$ and $h_{\alpha \beta }^{v}$ are the
viscous stress tensor and viscous molecular field respectively and
are defined by
\begin{eqnarray}\label{eq_def_visc_stress}
        \sigma _{\alpha \beta }^{v} &=& \beta _{1}Q_{\alpha \beta }Q_{\mu \nu }A_{\mu \nu }+\beta _{4}A_{\alpha \beta }+\beta _{5}Q_{\alpha \mu }A_{\mu \beta }
        \nonumber \\ & & +\beta _{6}Q_{\beta \mu }A_{\mu \alpha }
        +\frac{\mu_{2}N_{\alpha \beta }}{2}
         -\mu _{1}Q_{\alpha \mu }N_{\mu \beta }\nonumber \\& &~~~~~~~+ \mu _{1}Q_{\beta \mu }N_{\mu \alpha }
    \end{eqnarray}
\begin{equation}\label{eq_def_visc_field}
    h_{\alpha \beta }^{v}=-\frac{1}{2}\mu _{2}A_{\alpha \beta } - \mu _{1}N_{\alpha \beta }
\end{equation}
Here $\beta _{i},\:\mu _{i}$ are equivalent to the ELP viscosities,
$N_{\alpha \beta }$ is the co-rotational derivative, $N_{\alpha
\beta }=\dot{Q}_{\alpha \beta }-\varepsilon _{\alpha \mu \nu }\omega
_{\mu }Q_{\nu \beta } -\varepsilon _{\beta \mu \nu }\omega _{\mu
}Q_{\alpha \nu }$. $A_{\alpha \beta }=\frac{1}{2}\left( \partial
_{\alpha }u_{\beta }+\partial _{\beta }u_{\alpha } \right) $ and
$W_{\alpha \beta }=\frac{1}{2}\left(
\partial _{\alpha }u_{\beta }-\partial _{\beta }u_{\alpha } \right)
$ are the symmetric and anti-symmetric velocity gradient tensors
with the vorticity being $\omega _{\gamma }=\frac{1}{2}\varepsilon
_{\gamma \alpha \beta }W_{\alpha \beta }$. $\sigma _{\alpha \beta
}^{EM}$ is the stress tensor arising from externally applied
electromagnetic fields~\cite{Kloos:1999.3425}
\begin{eqnarray}\label{eq_def_EM_stress}
    \sigma_{\alpha \beta }^{EM}&=&\frac{1}{2}\left( H_{\alpha }B_{\beta }+H_{\beta }B_{\alpha }\right)
     - \frac{H_{\gamma }B_{\gamma }}{2}\delta _{\alpha \beta }
                                \nonumber \\& & +\frac{1}{2}\left( E_{\alpha }D_{\beta }+E_{\beta }D_{\alpha }\right) - \frac{E_{\gamma }D_{\gamma }}{2}\delta _{\alpha \beta }
\end{eqnarray}
where $\mathbi{E}~(\mathbi{H})$ is the electric (magnetic) field
strength, $\mathbi{D}$ the electric displacement vector and
$\mathbi{B}$ the magnetic flux density.

Direct calculation of the trace and off-diagonal elements of
Eq.~(\ref{eq_qs_order}) shows that lagrange multipliers are given by
$\lambda = \frac{1}{3}\left( h_{\gamma \gamma }-\frac{1}{2}\mu
_{2}A_{\gamma \gamma }\right) $ and $\lambda _{\gamma } =
\frac{1}{2}\varepsilon _{\alpha \beta \gamma }h_{\alpha \beta }$.
The term in $A_{\gamma \gamma }$ is included in order to correct the
small compressibility errors that arise in LB techniques when the
condition upon the Mach number (velocity to speed of sound ratio),
$M \equiv |\mathbi{u}|/  c_{s} \ll 1$ is violated.

Order of magnitude estimates and experiments both show the influence
of the moment of inertia to be negligible; we therefore set $J=0$ in
Eq.~(\ref{eq_qs_order}). Following~\cite{Care:2000.L665}, the
viscous stress tensor and the equation of motion
Eq.~(\ref{eq_qs_order}) are recast in a form more suitable for the
LB development,
\begin{eqnarray}\label{eq_new_visc_stress}
            \sigma _{\alpha \beta }^{v}&=&\beta _{1}Q_{\alpha \beta }Q_{\mu \nu }A_{\mu \nu }+\beta _{4}A_{\alpha \beta }+\beta _{5}Q_{\alpha \mu }A_{\mu \beta }
            \nonumber \\ & & +\beta _{6}Q_{\beta \mu }A_{\mu \alpha }
                                +\frac{\mu _{2}h_{\alpha \beta }}{2\mu _{1}} -\frac{\mu _{2}\lambda \delta _{\alpha \beta }}{2\mu _{1}}
                                - \frac{\mu _{2}\varepsilon _{\alpha \beta \gamma }\lambda _{\gamma }}{2\mu _{1}}
                                \nonumber \\ & &- \frac{\mu _{2}^{2}A_{\alpha \beta }}{4\mu _{1}} - Q_{\alpha \mu }h_{\mu \beta } +Q_{\alpha \mu }\varepsilon _{\mu \beta \gamma }\lambda _{\gamma }
                                +\frac{\mu _{2}Q_{\alpha \mu }A_{\mu \beta }}{2} \nonumber \\ & & + Q_{\beta \mu }h_{\mu \alpha } - Q_{\beta \mu }\varepsilon _{\mu \alpha \gamma }\lambda _{\gamma } - \frac{\mu _{2}Q_{\beta \mu }A_{\mu \alpha }}{2}
\end{eqnarray}
\begin{eqnarray}\label{eq_new_order_evolu}
    \dot{Q}_{\alpha \beta }&=&\frac{h_{\alpha \beta }}{\mu _{1}} - \frac{\lambda \delta _{\alpha \beta }}{\mu _{1}}-\frac{\varepsilon _{\alpha \beta \gamma }\lambda _{\gamma }}{\mu _{1}}
                        -\frac{\mu _{2}A_{\alpha \beta }}{2\mu _{1}}
    \nonumber \\ & &+\varepsilon _{\alpha \epsilon \lambda }\omega _{\epsilon }Q_{\lambda \beta }
             +\varepsilon _{\beta \epsilon \lambda }\omega _{\epsilon }Q_{\alpha \lambda }
\end{eqnarray}

The derivation of expressions for the molecular field and distortion
stress tensor follows the phenomenological approaches for the free
energy of liquid crystals~\cite{Barbero:2001}. The global free
energy density is considered to be a sum of contributions arising
from a number of different physical phenomena $\mathcal{F}_{Global}
= \int F_{Bulk} \; d\mathbi{r} + \int F_{Surface} \; d\mathbi{S}$.
The free energy densities have the form
\begin{eqnarray}\label{eq_energy_sum}
    F_{Bulk} & =&  F_{LdG} +  F_{Elastic} +  F_{Electric} \nonumber\\&&
    +  F_{Magnetic} +  F_{Flexo}
\end{eqnarray}
where
\begin{eqnarray}\label{eq_def_bulk_f1}
 F_{LdG}& =& F_{iso} + \frac{1}{2}\alpha_{F}Q_{\alpha \beta }Q_{\beta
\alpha } - \beta_{F}Q_{\alpha \beta }Q_{\beta \gamma }Q_{\gamma
\alpha } \nonumber\\&& + \gamma_{F}Q_{\alpha \beta }Q_{\beta \alpha
}Q_{\mu \nu }Q_{\nu \mu } \\ \label{eq_def_bulk_f2}
        F_{Elastic} &=&\frac{1}{2}L_{1}\partial _{\mu}Q_{\nu \gamma }\partial _{\mu}Q_{\nu \gamma } + \frac{1}{2}L_{2}\partial _{\mu}Q_{\nu \mu }\partial _{\gamma}Q_{\nu \gamma}
        \nonumber\\&& + \frac{1}{2}L_{3}\partial _{\mu}Q_{\nu \gamma }\partial _{\gamma}Q_{\nu \mu }
        + \frac{1}{2}L_{4}Q_{\mu \nu }\partial _{\mu}Q_{\gamma \tau }\partial _{\nu}Q_{\gamma \tau }
        \nonumber\\&&+ \frac{4\pi L_{1}}{P_{ch}}\varepsilon _{\mu \nu \gamma }Q_{\mu \tau }\partial _{\nu }Q_{\gamma \tau }
        \nonumber\\&&- \frac{4\pi L_{4}}{P_{ch}}\varepsilon _{\mu \nu \gamma }Q_{\mu \eta }Q_{\eta \tau }\partial _{\nu }Q_{\gamma \tau }
        \nonumber\\&&+ \frac{6 \pi^{2}}{P_{ch}^{2}} \left( L_{1}Q_{\mu \nu}Q_{\nu \mu} -L_{4}Q_{\mu \nu}Q_{\nu \tau}Q_{\tau \mu} \right) \\
        \label{eq_def_bulk_f3} F_{Electric} &=& -\frac{1}{3}\epsilon _{0}\epsilon _{a}^{max}E_{\alpha }Q_{\alpha \beta }E_{\beta
        } - \frac{1}{6}\epsilon_{a}\epsilon_{\gamma \gamma}E^{2}\\
        \label{eq_def_bulk_f4} F_{Magnetic} &=& -\frac{1}{3}\mu _{0}\chi _{a}^{max}H_{\alpha }Q_{\alpha \beta }H_{\beta }
        - \frac{1}{6}\mu_{0}\chi_{\gamma \gamma}H^{2} \\
        \label{eq_def_bulk_f5} F_{Flexo} &=&   -\xi _{1}E_{\alpha }\partial _{\gamma }Q_{\alpha \gamma } - \xi _{2}E_{\alpha }Q_{\alpha \gamma }\partial _{\mu }Q_{\gamma \mu } \\
        \label{eq_def_bulk_f6} F_{Surface} &=& \frac{W}{2}\left( Q_{\alpha \beta} - Q_{\alpha \beta}^{o} \right)^2
\end{eqnarray}

The coefficients $\alpha _{F}, \beta _{F}, \gamma _{F}$ are
parameters controlling the phase of the thermotropic liquid crystal,
the negative (positive) sign preceding the $\beta _{F}$ term
dictates a calamatic (discotic) state; for biaxial phases sixth
order terms are used. $L_{i},~i=1\dots 4$, determine the elastic
constants. $P_{ch}$ is the pitch of any chirality with
$\mu_{0}\left( \epsilon _{0}\right)$ being the
permeability(permittivity) of free space. \boldmath
$\chi$\unboldmath~ and \boldmath $\epsilon$ \unboldmath are the
diamagnetic and dielectric tensors with $\chi_{a}^{max}\left(
\epsilon _{a}^{max}\right)$ the maximal diamagnetic (dielectric)
 anisotropy (\textit{ie} $S=1$). $\xi _{1}$ and $\xi _{2} $ are flexoelectric
constants, $W$ an anchoring strength and $Q_{\alpha \beta }^{0}$ a
preferred surface state. This form for the free energy maintains
consistency with the Q-tensor dynamics
equations~\cite{Qian:1998.7475} in that a direct analogy with the
experimental ELP parameters is made (see \A \ref{parameters} for the
relation between experimental ELP values and the $\mathbi{Q}$-tensor
method).

To close the governing equations at surfaces, non-slip boundary
conditions are imposed upon the velocity.  For infinitely strong
anchoring a $\mathbi{Q}$ is specified according to
Eq.~(\ref{eq_q_tensor}). In cases of weak anchoring the order tensor
at the surface evolves according to
\begin{equation}\label{eq_surf_01}
    \mu_{S}\partial_{t}Q_{\alpha \beta} = h_{\alpha \beta }^{S} - \lambda^{S}\delta_{\alpha \beta}-\varepsilon_{\alpha \beta \gamma}\lambda_{\gamma}^{S}
\end{equation}
where $h_{\alpha \beta}^{S} = -\frac{\partial
F_{Bulk}}{\partial\left( \partial_{\tau}Q_{\alpha
\beta}\right)}\hat{\nu}_{\tau} - \frac{\partial
F_{Surface}}{\partial Q_{\alpha \beta}}$,
$\lambda^{S}=\frac{1}{3}h_{\gamma \gamma}^{S}$,
$\lambda_{\gamma}^{S}=\frac{1}{3}\varepsilon_{\alpha \beta
\gamma}h_{\alpha \beta}^{S}$, \boldmath $\hat{\nu}$ \unboldmath is
an outward pointing surface unit normal vector and $\mu_{S}$ is the
surface viscosity defined through $\mu_{S}=\mu_{1}l_{S}$ where
$l_{S}$ is a characteristic surface length typically being in the
range $l_{S}\approx 100\AA \rightarrow 1000\AA $
\cite{Vilfan:2001.061709}.

A non-dimensionalisation of the governing equations with respect to
characteristic velocity, $\bar{U}$, length , $\bar{L}$, viscosity,
$\eta_{eff}=\frac{1}{2}\left( \beta_{4}-\frac{\mu_{1}^{2}}{4\mu_{1}}
\right)$, and elastic constant, $\bar{L}_{1}$, yields three key
dimensionless numbers which govern the dynamics of the momentum,
director, and order parameter respectively
\begin{equation}\label{eq_si_non_dimen}
    \begin{array}{lll}
        Re = \frac{\rho \bar{U} \bar{L}}{\eta_{eff}}&,\;\;\;\;\;\;\; & \bar{\tau _{p}}= \frac{\rho \bar{L}^{2}}{\eta _{eff}}\\
        Er = \frac{\mu _{1} \bar{U} \bar{L}}{\bar{L}_{1}}&,\;\;\;\;\;\;\; & \bar{\tau _{n}}= \frac{\mu _{1} \bar{L}^{2}}{\bar{L}_{1}}\\
        De = \frac{\mu _{1} \bar{U}}{\alpha _{F} \bar{L}}&,\;\;\;\;\;\;\; & \bar{\tau _{s}}= \frac{\mu _{1}}{\alpha _{F}}
   \end{array}
\end{equation}
The characteristic timescales, $\bar{\tau}$, for variations in the
momentum, director and order parameter are also given. $Re$ and $Er$
are the Reynolds and Ericksen numbers. $De$ is the ratio of the
relaxation time for the order parameter, $\mu_{1}/\alpha_{F}$, to a
time scale associated with the flow, $\bar{L}/\bar{U}$; it is
similar to a Deborah number. Considering typical device parameters,
$\rho \sim 10^{3}\rm{kg\:m^{-3}}$, $\eta_{eff}\sim
10^{-2}\rm{kg\:m^{1}\:s^{-1}}$, $\bar{L}\sim 10^{-6}\rm{m}$,
$\bar{U}\sim 10^{-6}\rm{m\:s^{-1}}$, $\bar{L}_{1}\sim
10^{-12}\rm{kg\:m\:s^{-2}}$ and $\alpha_{F}\sim
10^{5}\rm{kg\:m^{-1}\:s^{-2}}$, we may estimate $\bar{\tau}_{p}\sim
10^{-7}$ s, $\bar{\tau}_{n}\sim 10^{-2}$ s, $\bar{\tau}_{s}\sim
10^{-7}$ s. It is apparent the relaxation rate of the momentum
compared to the director is much quicker, as is the relaxation of
the order compared to the director and accounting for these
timescale differences is essential for dynamic calculations.

\section{The Algorithm} \label{algorithm}

We proceed now to describe the LB method which recovers the set of
equations set out in Sec.~\ref{model}. The algorithm is defined in
Sec.~\ref{algorithm_statement}. In Sec.~\ref{algorithm_timescales}
we address the different timescales involved in liquid crystals'
dynamics and how to implement these in the LB method. A
Chapman-Enskog multi-scale analysis of the LB algorithm is given in
\A\ref{algorithm_analysis}.

\subsection{Statement Of The Algorithm} \label{algorithm_statement}

LBGK algorithms (\eg~\cite{Qian:1995.195}) are well established for
solving the Navier Stokes equations for isotropic
fluids~\cite{Dupin:2004.1885}. In order to recover the Qian-Sheng
equations of Sec~\ref{model} we introduce two LBGK algorithms, one
for the evolution of the momentum based on a scalar density
$f_{i}\left( \mathbi{x},t\right)$ and a second LBGK scheme based on
a tensor density $g_{i\alpha \beta }\left( \mathbi{x},t\right)$ to
recover the order tensor evolution.  It is important distinguish
between SI symbols in Sec.~\ref{model} and the symbols used in the
LB algorithms, which are defined in terms of lattice units. However
in this section, and Sec.~\A\ref{algorithm_analysis}, this
distinction is ignored for clarity. In
Sec.~\ref{algorithm_timescales} and \A\ref{parameters} the
distinction becomes important and a prime is used to denote a
lattice value. Further, a superscript $^{P}$ ($^{Q}$) is used to
distinguish between momentum (order) algorithms.

The principal reason for separating the momentum and order evolution
algorithms is the very large difference in time scales between the
two processes noted above.  In each algorithm, forcing terms are
used to recover the required additional terms in the stress tensor
and order evolution equations. This approach is more straightforward
to implement than the anisotropic scattering method used in an
earlier work~\cite{Care:2003.061703}.

The LBGK algorithm for an isotropic fluid may be written in the form
\begin{eqnarray}\label{eq_f_lbgk}
    \lefteqn{f_{i}\left( \mathbi{x}+\mathbi{c}_{i}\triangle t , t+\triangle t\right) = f_{i}\left( \mathbi{x},t\right)}
    \\&& ~~~~~~~-\frac{1}{\tau _{_{P}}}\left( f_{i}\left( \mathbi{x},t\right)-f_{i}^{(eq)}\left(
    \mathbi{x},t\right)\right)+ \phi_{i}\left( \mathbi{x},t\right) \nonumber
\end{eqnarray}
where $f_{i}\left( \mathbi{x},t\right)$ is the distribution function
for particles with velocity $\mathbi{c}_i$ at position $\mathbi{x}$
and time $t$, and $\triangle t$ is the time increment.
$f_{i}^{(eq)}\left( \mathbi{x},t\right)$ is the equilibrium
distribution function and $\tau _{_{P}}$ is the LBGK relaxation
parameter. The algorithm fluid density and velocity are determined
by the moments of the distribution function,
\begin{equation}\label{eq_f_lbgk_moments}
    \sum _{i} f_{i}
    \left[
    \begin{array}{c}
        1 \\
        c_{i\alpha }\\
    \end{array}
    \right]
    \;\;=\;
    \left[
    \begin{array}{c}
        \rho (\mathbi{x},t) \\
        \rho u_{\alpha }(\mathbi{x},t)\\
    \end{array}
    \right]
\end{equation}
The mesoscale equilibrium distribution function appropriate to
recover the correct hydrodynamics of incompressible fluids ($M \ll
1$) is,
\begin{equation}\label{eq_f_equilib}
    f_{i}^{(eq)}=t_{i}\rho \left[ 1+\frac{c_{i\alpha }u_{\alpha }}{c_{s}^{2}} + u_{\alpha }u_{\beta }\left( \frac{c_{i\alpha }c_{i\beta }-c_{s}^{2}\delta _{\alpha \beta }}{2c_{s}^{4}} \right) \right]
\end{equation}
where $t_{i}$ are lattice weights. $t_{i}$, $\mathbi{c}_{i}$,
$c_{s}$ are all dependant upon the choice of lattice, appropriate
values of these parameters are summarised in~\cite{Dupin:2004.1885}.
An analysis of the standard isotropic algorithm identifies the
lattice pressure and kinematic viscosity to be given by
\begin{equation}\label{eq_f_press_n_visc}
    P=\rho c_{s}^{2}\;\;\;\;\;\;\;\;\;\;\;\;\;\;\;\;\;\;\;\nu =\frac{c_{s}^{2}}{2}\left( 2\tau _{_{P}}-1
    \right)\triangle t
\end{equation}
$\phi _{i}$ is a forcing term which is chosen to recover the
required terms in the stress tensor. For a nematic liquid crystal
governed by Eq.~(\ref{eq_qs_momentum}) it is defined to be
\begin{equation}
\label{eq:phi} \phi _{i} = t_{i}c_{i\lambda }\partial _{\beta
}F_{\beta \lambda }
\end{equation}
where
\begin{eqnarray}\label{eq_lb_mom_foreterm}
F_{\alpha \beta } &=& \frac{\triangle t}{c_{s}^{2}}\left[
\sigma_{\alpha \beta }^{d} + \sigma_{\alpha \beta }^{EM} + \beta
_{1}Q_{\alpha \beta }Q_{\mu \nu }A_{\mu \nu } \right. \nonumber \\
&& +\beta _{5}Q_{\alpha \mu }A_{\mu \beta } +\beta _{6}Q_{\beta \mu
}A_{\mu \alpha } +\frac{\mu _{2}h_{\alpha \beta }}{2\mu _{1}}
\nonumber\\&& - \frac{\mu _{2}\varepsilon _{\alpha \beta \gamma
}\lambda _{\gamma }}{2\mu _{1}} - Q_{\alpha \mu }h_{\mu \beta }
+Q_{\alpha \mu }\varepsilon _{\mu \beta \gamma }\lambda _{\gamma }
\nonumber\\&&
 +\frac{\mu _{2}Q_{\alpha \mu }A_{\mu \beta
}}{2} + Q_{\beta \mu }h_{\mu \alpha } - Q_{\beta \mu }\varepsilon
_{\mu \alpha \gamma }\lambda _{\gamma } - \nonumber\\&& \left.
\frac{\mu _{2}Q_{\beta \mu }A_{\mu \alpha }}{2} \right]
    \end{eqnarray}
with analysis  identifying (see \A \ref{sec:derive:ce:mom})
\begin{equation}\label{eq_f_anals}
    P = \rho c_{s}^{2} + \frac{\mu _{2}\lambda }{2\mu _{1}}~;~~
    \rho c_{s}^{2}\left( 2\tau _{_{P}}-1 \right)\triangle t   =  \beta _{4}-\frac{\mu _{2}^{2}}{4\mu _{1}}
\end{equation}
and a macroscopic observable velocity of $\rho v_{\alpha} \equiv
\sum_{i}f_{i}c_{i\alpha}+(\triangle t/2)\partial_{\beta}F_{\beta
\alpha}$. The latter redefinition of the velocity is necessary to
reduce higher order artifacts which are introduced by a position
dependent forcing term~\cite{Guo:2002.046308}.

To recover the order evolution Eq.~(\ref{eq_new_order_evolu}) we
retain the simple LBGK form but replace the scalar density
$f_{i}\left( \mathbi{x},t\right)$ with a symmetric tensor
distribution $g_{i\alpha \beta }\left( \mathbi{x},t\right)$ evolving
according to
\begin{eqnarray}\label{eq_g_lbgk}
    \lefteqn{ g_{i\alpha \beta}\left( \mathbi{x}+\mathbi{c}_{i}\triangle t ,t+\triangle t\right)  =  g_{i \alpha \beta }\left(
    \mathbi{x},t\right)}
    \\ && - \frac{1}{\tau _{_{Q}}}\left( g_{i\alpha \beta }\left( \mathbi{x},t\right)-g_{i\alpha \beta }^{(eq)}\left(
     \mathbi{x},t\right)\right)+ \chi_{i\alpha \beta }\left( \mathbi{x},t\right) \nonumber
\end{eqnarray}
Here $g_{i\alpha \beta }^{(eq)}\left( \mathbi{x},t\right)$ is the
equilibrium order distribution function and $\tau _{_{Q}}$ the LBGK
relaxation parameter for the order evolution. The lowest moment of
the order distribution function, and its associated equilibrium
function, are defined to recover the order tensor of unit trace,
$S_{\alpha \beta }$
\begin{equation}\label{eq_g_moment}
     S_{\alpha \beta } = \sum_{i}g_{i\alpha \beta }=\sum_{i}g_{i\alpha \beta
    }^{(eq)}
\end{equation}
which is simply related to the dimensionless zero trace order
parameter $\mathbi{Q}$ through the relation
\begin{equation}\label{eq_s_to_q}
    Q_{\alpha \beta } = \frac{3S_{\alpha \beta }- \delta _{\alpha \beta
    }}{2}.
\end{equation}
The equilibrium order distribution is taken to be
\begin{eqnarray}\label{eq_g_equilib}
    g_{i\mu \nu }^{(eq)} & = & t_{i} S_{\mu \nu } \left[ 1+\frac{c_{i\alpha }u_{\alpha
    }}{c_{s}^{2}} \right.
    \nonumber \\  & & \left.~~~~+ u_{\alpha }u_{\beta }\left( \frac{c_{i\alpha }c_{i\beta }-c_{s}^{2}\delta _{\alpha \beta }}{2c_{s}^{4}} \right) \right]
\end{eqnarray}
$t_{i}, c_{i\alpha }, c_{s}^{2}$ are the same lattice parameters
defined for the momentum evolution. The forcing term  $\chi
_{i\alpha \beta }$ is chosen to provide the  rotational forces
required correctly to  recover Eq.~(\ref{eq_new_order_evolu})
\begin{equation}\label{eq_chi_force}
    \begin{array}{rc}
        \chi_{i\alpha \beta }=\frac{2 t_{i}\triangle t}{3} \left[ \frac{h_{\alpha \beta }}{\mu _{1}} - \frac{L_{1}\partial _{\lambda }\partial _{\lambda }Q_{\alpha \beta }}{\mu _{1}} -\frac{\lambda \delta _{\alpha \beta }}{\mu _{1}} -\frac{\varepsilon _{\alpha \beta \gamma }\lambda _{\gamma }}{\mu _{1}} \right.\\
                            \left. -\frac{\mu _{2}A_{\alpha \beta }}{2\mu _{1}} + \varepsilon _{\alpha \epsilon \lambda }\omega _{\epsilon }Q_{\lambda \beta }+\varepsilon _{\beta \epsilon \lambda }\omega _{\epsilon }Q_{\alpha \lambda } \right]
     \end{array}
\end{equation}
The analysis (see Sec.~\ref{sec:derive:ce:ord}) identifies the key
relation
\begin{equation}\label{eq_g_anals}
    \frac{c_{s}^{2}}{2}\left( 2\tau _{_{Q}}-1\right)\triangle t = \frac{L_{1}}{\mu _{1}}
\end{equation}
The scheme described here involves two coupled LB algorithms. Both
may be run independently; for example if the effect of flow is to be
ignored or only static equilibrium configurations are desired,
running the $g_{i\alpha \beta }$ scheme alone will suffice.  In
practice for typical device geometries, the flow fields evolve on a
much faster time scale than the director field; to model such
systems the momentum is evolved to steady state between each time
step of the order evolution equation. Although the time taken for
the momentum to reach equilibrium is significantly shorter than the
time step of the order evolution equation, the loss of accuracy in
this approach is small.

\subsection{Timescales In The Algorithm}
\label{algorithm_timescales}

Constructing an analogous set of dimensionless numbers to
Eq.~(\ref{eq_si_non_dimen}) in terms of the algorithm parameters,
from Eqs.~(\ref{eq:ch4:30}) and~(\ref{eq:ch4:46}) results in
\begin{equation}\label{eq_lb_non_dimen}
    \begin{array}{lll}
        Re^{\prime} = \frac{2\bar{U}^{\prime_{P}} \bar{L}^{\prime_{P}}}{c_{s}^{2}(2\tau_{_{P}}-1)}&,\;\;\;\;\;\;\; & \bar{\tau _{p}}^{\prime }= \frac{2\bar{L}^{\prime_{P} 2}}{c_{s}^{2}(2\tau _{_{P}}-1)}\\
        Er^{\prime} = \frac{2\bar{U}^{\prime_{Q}} \bar{L}^{\prime_{Q}}}{c_{s}^{2}(2\tau_{_{Q}}-1)}&,\;\;\;\;\;\;\; & \bar{\tau _{n}}^{\prime }= \frac{2\bar{L}^{\prime_{Q} 2}}{c_{s}^{2}(2\tau _{_{Q}}-1)}\\
        De^{\prime} = \frac{\mu _{1}^{\prime_{Q}} \bar{U}^{\prime_{Q}}}{\alpha _{F}^{\prime_{Q}} \bar{L}^{\prime_{Q}}}&,\;\;\;\;\;\;\; & \bar{\tau _{s}}^{\prime}= \frac{\mu _{1}^{\prime}}{\alpha _{F}^{\prime_{Q}}}
   \end{array}
\end{equation}
We choose $\tau_{_{P}}=\tau_{_{Q}}=1$ and
$\bar{L}^{\prime_{P}}=\bar{L}^{\prime_{Q}}\sim
\bar{L}\sqrt{\frac{\left.2\frac{\partial ^{2}F_{LdG}}{\partial
S^{2}}\right|_{S_{0}}}{3\bar{L}_{1}} }$. The latter identity sets
the simulation size to resolve variations in $\mathbi{Q}$.  The
correct dynamics are achieved by matching the algorithm
dimensionless numbers Eq.~(\ref{eq_lb_non_dimen}) to the real
dimensionless numbers Eq.~(\ref{eq_si_non_dimen}). From
Eq.~(\ref{eq_lb_non_dimen}) this requires that
$\bar{U}^{\prime_{P}}$ differs from $\bar{U}^{\prime_{Q}}$ by an
amount $Er/Re$. In order to recover an internally consistent
simulation, these different values of the LB velocities in the
momentum and order evolution algorithms require the forces to be
appropriately scaled when information is passed between the two
algorithms. A list of the scaling is given in \A\ref{parameters}.

We may now take the ratio of characteristic SI to LB times to give
the time value of the LB discrete time step. Using typical values
shows that: $\triangle t_{_{P}}\sim 10^{-13}$ s, $\triangle
t_{n}\sim 10^{-8}$ s, $\triangle t_{_{S}}\sim 10^{-8}$ s. Hence the
momentum algorithm needs to be iterated many times within a single
iteration of the order algorithm. Alternatively for laminar creeping
flows, $Re\ll 1$, the equilibrium flow field will be reached in a
small number of $\triangle t_{_{P}}$ and we may jump forward in time
to the next $\triangle t_{n} \sim \triangle t_{_{S}}$ reducing the
overall processing time.


\section{Results}
\label{results} In this section we begin by presenting results which
validate the algorithm developed above by comparing its numerical
predictions with analytical results for some simple cases. We then
show how the technique may be used to study the motion of defects in
a commonly studied bistable liquid crystal device.

\subsection{Comparison with analytical results for the Miesowicz viscosities}
\label{anal_results}

We first consider the flow alignment of the director in a shear flow
in the absence of an external aligning field. Provided the channel
width is sufficiently large, we may ignore gradients in $Q$ in the
centre of the channel. In this case the alignment at the centre of
the channel is solely determined by the viscous torque. From
Eqs.(\ref{eq_q_tensor}) and (\ref{eq_new_order_evolu}) we can solve
for the director angle, $\theta$, to find
\begin{equation}\label{eq_res_shear_angle}
    \cos (2\theta )=-\frac{\mu _{1}}{\mu _{2}}(3S+P_{_{B}})=
    -\frac{\gamma _{1}}{\gamma _{2}}\left( \frac{S+\frac{1}{3}P_{_{B}}}{S_{0}} \right)
\end{equation}

A second standard case is to measure the shear viscosity of the
nematic in the presence of a strong external field which imposes a
fixed director angle.  These experiments yield the Miesowicz
viscosities \cite{deGennes:1993}. However, the standard results must
be extended for the case of a variable order parameter. For an
arbitrary fixed director angle the effective viscosity,
$\eta^{\star}$, is found to be
\begin{equation}\label{eq:ch4:74}
    \begin{array}{lllllllll}
    \eta^{\star} &= \frac{\sigma_{\alpha \beta}^{v}}{2A_{\alpha \beta}}  \\
         &= \frac{\beta_{4}}{2}+\frac{\mu_{2}}{8}S(3n_{1}^{2}-1)-\frac{\mu_{2}}{8}S(3n_{3}^{2}-1)\\&+\frac{\beta_{5}}{4}S(3n_{3}^{2}-1)
                    +\frac{9\beta_{1}}{4}S^{2}n_{1}^{2}n_{3}^{2} +\frac{\beta_{6}}{4}S(3n_{1}^{2}-1)\\ &+\frac{9\mu_{1}}{2}S^{2}n_{1}^{2}n_{3}^{2}
                    +\frac{9\mu_{1}}{8}S^{2}n_{2}^{2}n_{3}^{2} -\frac{9\mu_{1}}{8}S^{2}n_{1}^{2}n_{2}^{2}\\ &-\frac{\mu_{1}}{4}S^{2}(3n_{1}^{2}-1)(3n_{3}^{2}-1)+\frac{\mu_{1}}{8}S^{2}(3n_{3}^{2}-1)^{2}
                    \\ &+\frac{\mu_{1}}{8}S^{2}(3n_{1}^{2}-1)^{2}
    \end{array}
\end{equation}
from which the following Miesowicz viscosities can be determined:-
\begin{equation}\label{eq:ch4:73}
    \left.
    \begin{array}{llll}
        \eta_{a}=\frac{\beta_{4}}{2}-\frac{\beta_{5}S}{4}-\frac{\beta_{6}S}{4}&\rm{at}\;\hat{n}_{\alpha}=(0,1,0)\\
        \eta_{b}=\frac{\beta_{4}}{2}+\frac{3\mu_{2}S}{8}-\frac{\beta_{5}S}{4}+\frac{\beta_{6}S}{2}
        +\frac{9S^{2}\mu_{1}}{8}&\rm{at}\;\hat{n}_{\alpha}=(1,0,0)\\
        \eta_{c}=\frac{\beta_{4}}{2}-\frac{3\mu_{2}S}{8}+\frac{\beta_{5}S}{4}
        -\frac{\beta_{6}S}{2}+\frac{9S^{2}\mu_{1}}{8}&\rm{at}\;\hat{n}_{\alpha}=(0,0,1)
    \end{array}
    \right\}
\end{equation}
these being identical to the EL expressions~\cite{deGennes:1993} in
the limit $S\rightarrow S_{0}$. A biaxial correction is not required
as the aligning field serves to cancel biaxial contributions from
the shear.

In order to assess the accuracy of the method described in
\s\ref{algorithm} we tested it against these analytical values. We
used a channel width $L=1.2\mu \rm{m}$, a shear rate
$\dot{\gamma}=10^{4}\rm{s}^{-1}$, viscosities
\{$\alpha_{1}=-0.011,\:\alpha_{2}=-0.102,\:\alpha_{3}=-0.005,\:\alpha_{4}=0.074,\:\alpha_{5}=0.084,\:\alpha_{6}=-0.023$\}
$\rm{kg}$ $\rm{m^{-1}}$ $\rm{s^{-1}}$, Landau parameters \{
$a=65000\:\rm{J\:m^{-3}\: K^{-1}}$, $B=530000\:\rm{J\:m^{-3}}$,
$C=980000\:\rm{J\:m^{-3}}$\} and $T=T_{IN}-4(T_{IN}-T^{\star})$. The
boundaries were assumed to have infinite anchoring and the flow
induced by adding $2t_{i}\rho^{w} \underline{c}_{i}\cdot
\underline{u}^{w}/(c_{s}^{2})$ to the right hand side of
Eq.~(\ref{eq_f_lbgk}) where the wall velocity is
$+(-)\underline{u}^{w}$ at the top (bottom) boundaries with
periodicity in the $x$ and $y$ directions.

In the absence of an aligning field, the director angle in the shear
flow was found to be $12.166^{\circ}$ which agreed with the value
predicted by Eq.~(\ref{eq_res_shear_angle}) to 7 significant
figures. Accuracy was found to be maintained over all flow aligning
viscosity ratio's with a typical increase in $S$ around $0.002$ and
biaxiality $P_{_{B}}=0.002$. The Miesowicz viscosities were measured
using an aligning field of $75$ volts ($\triangle
\epsilon_{a}=10.3$) in the relevant directions.  Non-slip boundary
conditions were applied using the bounce-back
method~\cite{Dupin:2004.1885})  and the flow was induced by applying
a constant body force at $z=L/2$. The resultant viscosity ratio's
are compared in Table~\ref{tab:ch4_mies1} where data is measured at
$z=L/4$. It can be seen that the LB solver gives results in good
agreement with the expected values.
\begin{table}[!h]
\begin{center}
\begin{tabular}{||c|c|c||}
    \hline \hline
     Theory & Simulation & \% error \\
    \hline
     $\eta_{a}/\eta_{b}=1.446$ & $ \dot{\gamma}_{a}^{-1}/\dot{\gamma}_{b}^{-1}=1.446 $ & $1.6\times 10^{-4}$ \\
    \hline
     $\eta_{a}/\eta_{c}=0.227$ & $ \dot{\gamma}_{a}^{-1}/\dot{\gamma}_{c}^{-1}=0.227 $ & $1.4\times 10^{-4}$ \\
    \hline
     $\eta_{b}/\eta_{c}=0.157$ & $ \dot{\gamma}_{b}^{-1}/\dot{\gamma}_{c}^{-1}=0.157 $ & $2.8\times 10^{-5}$ \\
    \hline
    \hline
\end{tabular}
\caption{Table of theoretical (Eq.~(\ref{eq:ch4:73})) and simulated
ratio's of the Miesowicz viscosities.} \label{tab:ch4_mies1}
\end{center}
\end{table}


\subsection{Investigation of defect motion in a bistable device}
\label{device_results}

\begin{figure}[!t]
\begin{center}
\includegraphics[scale=0.5,angle=-0]{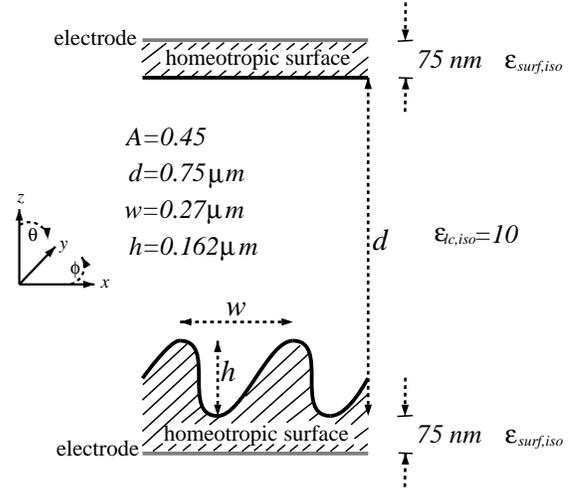}
\caption{\small{Schematic of the two dimensional ZBD geometry over
two grating pitches, $w$. Homeotropic boundary conditions serve to
cause bistability. Simulations contain one grating pitch and
periodic boundaries in the $x$ and $y$ directions.}} \label{fig:85}
\end{center}
\end{figure}

The ZBD device~\cite{patent, 181} uses a structured surface, such as
that in Fig.~(\ref{fig:85}), to introduce bistability which may be
used to design a very low power display. The two bistable states are
characterised by the presence or absence of defects which will be
referred to as the defect ($\mathcal{D}$) and continuous
($\mathcal{C}$) states respectively. One possible method to latch
switch between states is to use an electric field that couples to
the flexoelectric properties of the liquid crystal material.
Latching between the two states is a dynamic process which involves
the nucleation and annihilation of defects.

\begin{figure*}[!htb]
\begin{center}
\vspace{-0.5cm}
\includegraphics[scale=0.45,angle=-90]{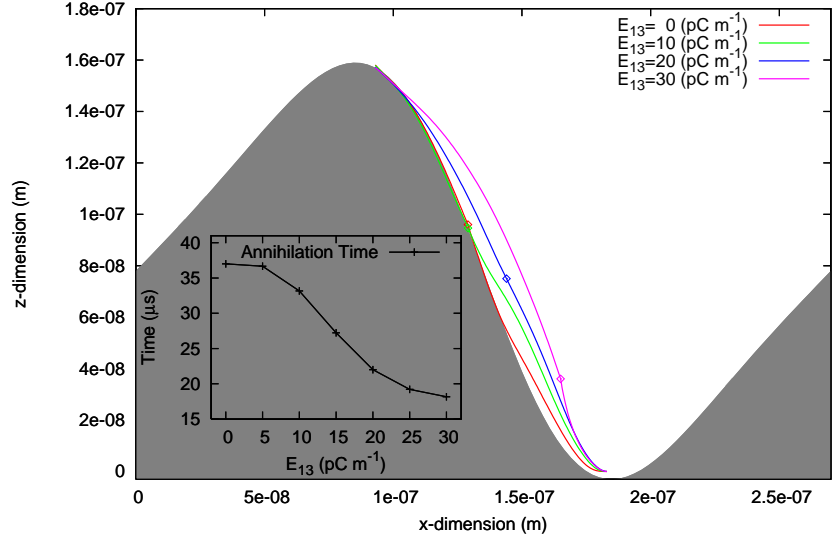}

\caption{\small{Defect trajectories during the $\mathcal{D}$ to
$\mathcal{C}$ latching for various $E_{13}$ values. Points indicate
location of the annihilation. The inset figure indicates the time at
which defects annihilate from the turn on of the voltage.
$\epsilon_{\gamma \gamma }=18$, $V=+18$ volts and $\triangle
\epsilon_{a}=10.3$.}} \label{fig:62}
\end{center}
\end{figure*}

 We followed
~\cite{patent} and modelled  the ZBD surface with the function
$g(x)=h\sin \left[ \frac{2\pi x}{w}+A\sin \left( \frac{2\pi x}{w}
\right)  \right]$ projected onto the LB boundary over one grating
period, $w$; the height of the grating is $h$ and the parameter $A$
controls the level of asymmetry. Weak anchoring conditions was used
at the surfaces by implementing Eq.~(\ref{eq_surf_01}) with an
explicit forward time finite difference method.  The gradients in
the equation were extrapolated to second order from the bulk and an
average taken over the values obtained from each lattice direction.
It should be noted that in order to achieve equality of the elastic
constants in the bulk and at the surface it is necessary for the
surface parameters $L_{i}^{S}$ to be different from bulk LB through
the $L_{i}^{S}=L_{i}/(2\tau_{_{Q}}-1)$. This arises because the
relaxation processes in the bulk, which are governed by the
parameter $\tau_{_{Q}}$, contribute to the measured elastic
constants in the bulk.  However, there is no equivalent collision
process in the surface algorithm.

In the presence of the voltage applied to the device, it is
necessary to solve Maxwell's equations over the LB grid to obtain
the local values for the electric field, $\mathbi{E}$.  For
completeness these equations are
\begin{equation}\label{eq_efiled_a}
\left.
\begin{array}{llllllllll}
    \partial _{\alpha }D_{\alpha } = 0 \\
    D_{\alpha } = \epsilon _{0}\epsilon _{\alpha \beta }E_{\beta }+P_{\alpha } \\
    E_{\beta } =-\partial _{\beta }V \\
    \epsilon _{\alpha \beta }= (2\triangle \epsilon _{a}^{max}Q_{\alpha \beta }+\epsilon _{\gamma \gamma }\delta _{\alpha \beta })/3 \\
\end{array}
\right\}
\end{equation}
in which $V$ is the local voltage, $\epsilon_{\gamma
\gamma}=2\epsilon_{\perp}+\epsilon_{\parallel}$, $\triangle
\epsilon_{a}=S\triangle \epsilon_{a}^{max}$ and $P_{\sigma }$ is
defined from Eq.~(\ref{eq_def_bulk_f5}) by writing it in the form
$F_{Flexo}=-P_{\sigma}E_{\sigma}$. We solve equations
(\ref{eq_efiled_a}) using a successive overrelaxation method at each
iteration of the LB algorithm for $g_{i\alpha \beta}$.  This
therefore determines the electric field which is consistent with the
instantaneous value of the $Q$ tensor.

We investigated the effect of material properties on the motion of
defects in this device; in particular we studied the interplay of
dielectric, flexoelectric and surface polarisation effects. We used
the set of material parameters given in \A\ref{parameters}. The
system was first established at steady state in one of the
equilibrium states; the simulation was then run using the algorithm
described above.  The equilibrium states were located by starting
from an appropriate initial condition and running only the $g_{i
\alpha\beta}$ algorithm.  The defect equilibrium state
($\mathcal{D}$) has a -1/2 defect near the peak of the grating and a
+1/2 defect near the trough of the grating.

\begin{figure*}[!htb]
\begin{center}
\includegraphics[scale=0.45,angle=-90]{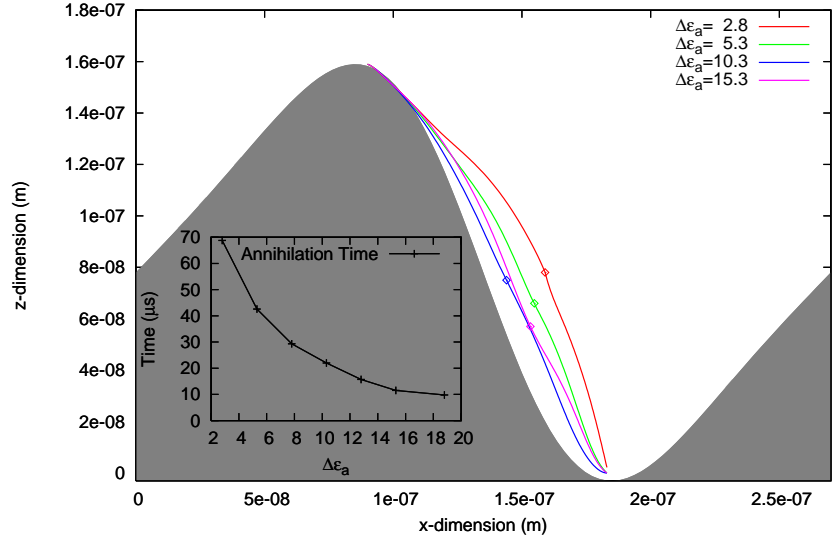}
\caption{\small{Defect trajectories during the $\mathcal{D}$ to
$\mathcal{C}$ latching for various $\triangle \epsilon_{a}$ values.
Points indicate location of the annihilation. The inset figure
indicates the time at which defects annihilate from the turn on of
the voltage. $\epsilon_{\gamma \gamma }=18$, $V=+18$ volts and
$E_{13}=20 \rm{pC\;m^{-1}}$.}} \label{fig:63}
\end{center}
\end{figure*}

In Fig.~(\ref{fig:62}) the flexoelectric coefficient
$E_{13}=\frac{e_{11}+e_{33}}{2}$, ($e_{11}=e_{33}$) is varied.
Starting in the $\mathcal{D}$ state and applying $+18$($0$) volts to
the upper (lower) electrodes the resultant defect trajectories are
shown in the grating region. For $E_{13}=0$ the defects move slowly
along the surface and annihilate. As $E_{13}$ increases we increase
the surface polarisation and order parameter which pushes the
defects further out into the bulk of the device to annihilate. As
$E_{13}$ increases the $-\frac{1}{2}$ defect mobility is increased
as seen in the annihilation locations. The inset of
Fig.~(\ref{fig:62}) shows the time taken for the defects to
annihilate which is an indicator of the latching speed. It is found
increasing $E_{13}$ increases the latching speed.

Alternatively we may keep $E_{13}$ constant and vary the dielectric
anisotropy, see Fig.~(\ref{fig:63}). For a deceasing $\triangle
\epsilon_{a}$ we effectively increase flexoelectric contributions to
the nematic (see Eqs.~(\ref{eq_def_bulk_f3})and
(\ref{eq_def_bulk_f5})), this increases the surface polarisation
that pushes the defects further away from the surface. Increasing
$\triangle \epsilon_{a}$ effectively reduces the flexoelectric
contributions and the defects annihilate closer to the surface; this
also reduces the latching time.

\begin{figure*}[!htb]
\begin{center}
\vspace{-0.5cm}
\includegraphics[scale=0.45,angle=-90]{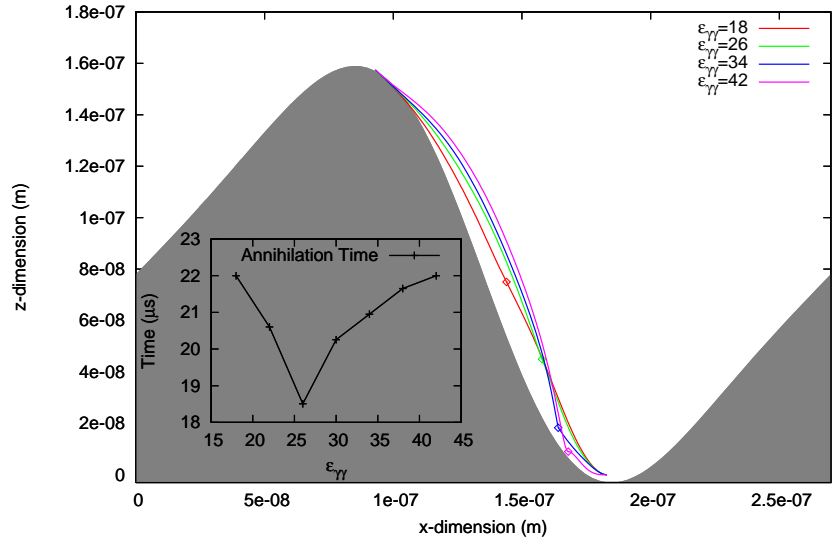}
\caption{\small{Defect trajectories during the $\mathcal{D}$ to
$\mathcal{C}$ latching for various $\epsilon_{\gamma \gamma }^{S}$
values. Points indicate location of the annihilation. The inset
figure indicates the time at which defects annihilate from the turn
on of the voltage. $\triangle \epsilon_{a}=10.3$, $V=+18$ volts and
$E_{13}=20 \rm{pC\;m^{-1}}$.}} \label{fig:64}
\end{center}
\end{figure*}

Fig.~(\ref{fig:64}) has fixed $\triangle \epsilon_{a}$ and $E_{13}$
but the grating permittivity $\epsilon_{\gamma \gamma }^{S}$ is
changed. This has the effect of diffracting the electric field lines
for an increased mis-match of surface and nematic permittivities. At
the lower dielectric mismatch the defect annihilation location is at
$\sim h/2$. Increasing the dielectric mismatch increases the
mobility of the $-\frac{1}{2}$ defect allowing it to travel further
and annihilate near the grating trough. There appears to be an
optimum value of $\epsilon_{\gamma \gamma }^{S} \sim 26 $ for which
the annihilation time  is shortest.
\begin{figure*}[!htb]
\begin{center}
\includegraphics[scale=0.45,angle=-90]{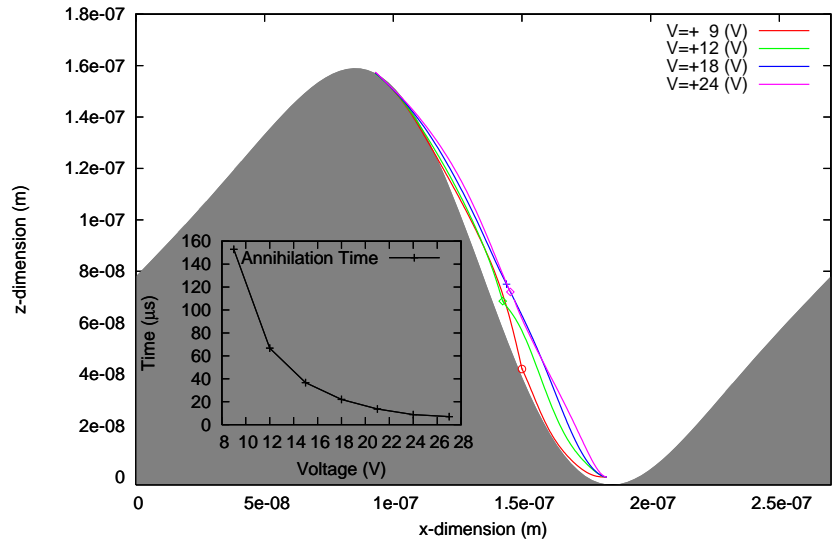}
\caption{\small{Defect trajectories during the $\mathcal{D}$ to
$\mathcal{C}$ latching for various $E_{13}$ values. Points indicate
location of the annihilation. The inset figure indicates the time at
which defects annihilate from the turn on of the voltage. $\triangle
\epsilon_{a}=10.3$, $\epsilon_{\gamma \gamma }=18$ and $E_{13}=20
\rm{pC\;m^{-1}}$.}} \label{fig:65}
\end{center}
\end{figure*}

Fig.~(\ref{fig:65}) shows the effect of increasing the applied
voltage for constant $\epsilon_{\gamma \gamma }^{S}$, $\triangle
\epsilon_{a}$ and $E_{13}$ ; this has the effect of increasing the
contributions of both the flexoelectric and dielectric terms. An
increased voltage tends to cause the defect trajectories to move
away from the surface towards a saturation distance for which
further increase causes little difference. Above the voltages shown
in the figure, a different latching mode is seen in which several
pairs of defects occur in the annihilation process. As with a
Fredericksz response an increased voltage results in a faster
latching response.


\section{Conclusions}
\label{conclusions}

An LB method has been presented which can be used to predict the
dynamics of a nematic liquid crystal in the complex geometries which
are increasingly being adopted for display devices. Nematic order,
director, velocity, electric fields and surface polarisations are
all recovered; this allows comparison to experimental results to be
made for a wide range of cell geometries or surface patterning. In
the presence of structured surfaces and defects, it is essential to
consider the variation in the order parameter. Essentially, a full
`device solver' has been developed and example results are given
that show both the accuracy of the solver and its use in determining
behaviour of next generation LC devices. The influence of surface
polarisations resulting from dielectric and flexoelectric effects
are shown to effect defect trajectories and ultimately latching
speeds. The solver is currently being used in the development of
next generation bistable devices.


\acknowledgements \label{acknowledge} We thank Dr. I.Halliday, Dr.
D.J. Cleaver, Dr. S.V. Lishchuk,  from Sheffield Hallam University
and Dr. J.C. Jones, Dr. R. Amos from ZBD Displays Ltd for very
useful discussions and comments in the developments and
investigations undertaken in this paper.



\appendix

\section{Chapman-Enskog analysis of the LB algorithm}


\label{algorithm_analysis} In this section we present a
Chapman-Enskog analysis of the momentum and order evolution schemes.
This analysis serves two purposes: to demonstrate that the method
recovers the required governing equations and to identify the
relation of the LBGK parameters to the associated transport
coefficients and forcing terms.

\subsection{Momentum Evolution} \label{sec:derive:ce:mom}  The
moments of the distribution function, $f_i$, are defined to be
\begin{equation}\label{eq:ch4:20a}
    \left.
    \begin{array}{lllllllll}
        \sum_{i}f_{i}^{(0)}
            \left[
            \begin{array}{ccccc}
                1 \\
                c_{i\alpha} \\
                c_{i\alpha }c_{i \beta}
            \end{array}
            \right] &=& \left[
            \begin{array}{ccccc}
                \rho \\
                \rho u_{\alpha} \\
                \Pi_{\alpha \beta}^{(0)}=\rho c_{s}^{2}\delta_{\alpha \beta}+\rho u_{\alpha}u_{\beta}
            \end{array}
            \right]
            \\
            \sum_{i}f_{i}^{(n)}
            \left[
            \begin{array}{ccccc}
                1 \\
                c_{i\alpha} \\
                c_{i\alpha }c_{i\beta}
            \end{array}
            \right] &=& \left[
            \begin{array}{ccccc}
                0 \\
                0 \\
                \Pi_{\alpha \beta}^{(n)}
            \end{array}
            \right] \;\;,\;\;\;\; n > 0
    \end{array}
    \right\}
\end{equation}
The the velocity basis and the $t_{i}$ are chosen to give
\begin{equation}\label{eq:ch4:20b}
    \left.
    \begin{array}{lll}
       \sum_{i}t_{i}&=&1 \\
        \sum_{i}t_{i}c_{i\alpha}&=&0 \\
       \sum_{i}t_{i}c_{i\alpha}c_{i\beta}&=&c_{s}^{2}\delta_{\alpha \beta} \\
       \sum_{i}t_{i}c_{i\alpha}c_{i\beta}c_{i\gamma}&=&0 \\
      \sum_{i}t_{i}c_{i\alpha}c_{i\beta}c_{i\gamma}c_{i\theta}&=&c_{s}^{4} \Delta_{\alpha \beta \gamma \theta}
   \end{array}
    \right\}
\end{equation}
where
\begin{equation}
 \Delta_{\alpha \beta \gamma \theta} =  \delta_{\alpha \beta}\delta_{\gamma \theta}+\delta_{\alpha \gamma}\delta_{\beta \theta}+\delta_{\alpha \theta}\delta_{\beta \gamma}
\end{equation}
Using a Taylor expansion on the left hand side of
Eq.~(\ref{eq_f_lbgk}) we obtain:
\begin{equation}\label{eq:ch4:21}
    \begin{array}{ccc}
        \triangle t \partial_{t}f_{i}+\frac{\triangle t^{2}}{2}\partial_{t}\partial_{t}f_{i} +
        \triangle t c_{i{\alpha}}\partial_{\alpha}f_{i} + \triangle t^{2}c_{i\alpha}\partial_{t}\partial_{\alpha}f_{i}\\
        +\frac{\triangle t^{2}}{2}c_{i\alpha }c_{i\beta}\partial_{\alpha} \partial_{\beta}f_{i}
        = -\frac{1}{\tau_{_{P}}}\left( f_{i}-f_{i}^{(eq)} \right) + \phi_{i}
    \end{array}
\end{equation}
We assume a forcing term $\phi_{i}$ of the form
$\phi_{i}=t_{i}c_{i\lambda} \partial_{\beta}F_{\beta \lambda}$ and
use a multi-scale expansion, to second order
\begin{equation}\label{eq:ch4:22}
    \left.
    \begin{array}{ccc}
        t_{1}=\varepsilon t\;\;,\;\;\;\;\;t_{2}=\varepsilon^{2}t\;\;,\;\;\;\;\;\partial_{t}=\varepsilon \partial_{t_{1}} +\varepsilon^{2}\partial_{t_{2}} \\
        x_{1}=\varepsilon x\;\;,\;\;\;\;\;x_{2}=\varepsilon^{2}x\;\;,\;\;\;\;\;\partial_{x}=\varepsilon \partial_{x_{1}} +\varepsilon^{2}\partial_{x_{2}} \\
        f_{i}=f_{i}^{(0)}+\varepsilon f_{i}^{(1)}+\varepsilon^{2}f_{i}^{(2)}
    \end{array}
    \right\}
\end{equation}
Using this expansion in \e(\ref{eq:ch4:21}) and collecting terms we
obtain:
\newline
\noindent $O(\varepsilon ^{0})$
\begin{equation}
f_{i}^{(0)} = f_{i}^{(eq)} \label{eq:ch4:24a}
\end{equation}
$ O(\varepsilon^{1})$
\begin{eqnarray}
-\tau_{_{P}}\triangle t (\partial_{t_{1}}+
c_{i\alpha_{1}}\partial_{\alpha_{1}} )f_{i}^{(0)}\nonumber \\+
\tau_{_{P}}t_{i}c_{i\lambda}\partial_{\beta_{1}}F_{\beta \lambda}
 &=&  f_{i}^{(1)} \label{eq:ch4:24b}
\end{eqnarray}
$O(\varepsilon ^{2})$
\begin{eqnarray}
 \left(\frac{1}{2}-\tau_{_{P}}\right) \triangle t\left( \partial_{t_{1}}+c_{i\alpha}\partial_{\alpha_{1}}
 \right)f_{i}^{(1)}\nonumber\\
- \tau_{_{P}}\triangle t \left(
\partial_{t_{2}}+c_{i\alpha}\partial_{\alpha_{2}} \right)
f_{i}^{(0)} \nonumber\\
 -\frac{\tau_{_{P}}\triangle t }{2}\left( \partial_{t_{1}}+c_{i\alpha}\partial_{\alpha_{1}} \right) t_{i}c_{i\lambda}\partial_{\beta_{1}}F_{\beta \lambda }\nonumber \\+\tau_{_{P}}t_{i}c_{i\lambda}\partial_{\beta_{2}}F_{\beta \lambda }&=& f_{i}^{(2)}
 \label{eq:ch4:24c}
\end{eqnarray}
in which we have used the $O(\varepsilon ^{1})$ result of
\e(\ref{eq:ch4:24b}) to replace a term of the form
$(\partial_{t_{1}}+c_{i\alpha}\partial_{\alpha _{1}})f_{i}^{(0)}$ in
the $O(\varepsilon ^{2})$ result. Taking the zeroth moment of
\e(\ref{eq:ch4:24b}) and \e(\ref{eq:ch4:24c}) whilst respecting
\e(\ref{eq:ch4:20a}) yields:
\begin{equation}\label{eq:ch4:25}
    \begin{array}{lllll}
        O(\varepsilon ^{1}) & &  &\partial_{t_{1}}\rho + \partial_{\alpha_{1}}\left( \rho u_{\alpha} \right) = 0 \\
        O(\varepsilon ^{2}) & &  &\partial_{t_{2}}\rho + \partial_{\alpha_{2}}\left( \rho u_{\alpha} \right) = c_{s}^{2}\partial_{\gamma_{1}}\partial_{\beta_{1}}F_{\beta
        \gamma}/2
    \end{array}
\end{equation}
which can be recombined to give the continuity equation:
\begin{equation}\label{eq:ch4:25aa}
    \partial_{t}\rho + \partial_{\alpha}\left( \rho u_{\alpha} \right) = 0
\end{equation}
where the term  $c_{s}^{2} \partial_{\lambda}
\partial_{\beta}F_{\beta \lambda}/2$ is corrected for by redefining the macroscopic velocity as described
below Eq.~(\ref{eq_f_anals}).


Taking the first moment ($\sum_{i}c_{i\beta}$) of the first and
second order \e(\ref{eq:ch4:24b}) and \e(\ref{eq:ch4:24c}) whilst
respecting \e(\ref{eq:ch4:20a}) yields:
\begin{equation}\label{eq:ch4:26}
    \begin{array}{lllll}
        O(\varepsilon ^{1}) & &  &\partial_{t_{1}}(\rho u_{\beta}) + \partial_{\alpha_{1}}\Pi_{\alpha \beta}^{(0)} = \frac{c_{s}^{2}}{\triangle t} \partial_{\gamma_{1}}F_{\gamma \beta} \\
        O(\varepsilon ^{2}) & &  &\left( 1-\frac{1}{2\tau_{_{P}}}  \right) \partial_{\alpha_{1}}\Pi_{\alpha \beta}^{(1)} + \partial_{t_{2}}(\rho u_{\beta}) \\&&&+ \partial_{\alpha_{2}}\Pi_{\alpha \beta}^{(0)} = \frac{c_{s}^{2}}{\triangle t}\partial_{\gamma_{2}}F_{\gamma \beta}
                                    - \frac{c_{s}^{2}}{2}\partial_{t_{1}}\partial_{\gamma_{1}}F_{\gamma \beta}
    \end{array}
\end{equation}
In order to progress, the $\Pi_{\alpha \beta}^{(1)}$ term needs
evaluating and this requires knowledge of $f_{i}^{(1)}$ (in
\e(\ref{eq:ch4:24b})). Using \e(\ref{eq_f_equilib}) to
$O(\underline{\mathbi{u}})$ in \e(\ref{eq:ch4:24b}), taking its
zeroth moment and back substituting the result
($\partial_{t_{1}}\rho = -
\partial_{\beta_{1}}(\rho u_{\beta})$), followed by taking the first
moment and another back substitution ($\partial_{t_{1}}(\rho
u_{\beta})= - c_{s}^{2}\partial_{\beta_{1}}\rho -
\tau_{_{P}}c_{s}^{2}\partial_{\gamma_{1}}F_{\gamma \beta}$) yields
\begin{eqnarray}\label{eq:ch4:27}
    f_{i}^{(1)} &=& -\frac{\tau_{_{P}}\triangle t \partial_{\beta _{1}}(\rho u_{\alpha})H_{i\alpha \beta}}{c_{s}^{2}}\nonumber\\
    &&~~~~+\tau_{_{P}}t_{i}c_{i\lambda}(\tau_{_{P}}\triangle t+1)\partial_{\gamma_{1}}F_{\gamma \lambda}
\end{eqnarray}
in which the symmetric quantity $H_{i\alpha \beta}$ is given by
\begin{equation}\label{eq:ch4:28}
    H_{i\alpha \beta} = t_{i}c_{s}^{2}\left( \frac{c_{i\alpha}c_{i\beta}}{c_{s}^{2}}-\delta_{\alpha \beta}  \right)
\end{equation}
The symmetry of \e(\ref{eq:ch4:28}) allowing us to replace
$\partial_{\lambda}(\rho u_{\beta})$ by $\rho A_{\lambda \beta}$ in
Eq. (\ref{eq:ch4:27}), in the incompressible limit.

We now use \e(\ref{eq:ch4:27}) to evaluate \es(\ref{eq:ch4:26}) and
combine the results to find
\begin{eqnarray}\label{eq:ch4:30}
    \lefteqn{\partial_{t}(\rho u_{\beta})+u_{\alpha}\partial_{\alpha}(\rho u_{\beta})= -\partial_{\beta}(\rho
c_{s}^{2})}\nonumber\\
&&+c_{s}^{2}(2\tau_{_{P}}-1)\triangle t\partial_{\alpha}(\rho
A_{\beta \alpha}) +\frac{c_{s}^{2}}{\triangle
t}\partial_{\alpha}F_{\alpha \beta }
\end{eqnarray}
Here the term
$-\frac{c_{s}^{2}}{2}\partial_{t}\partial_{\gamma}F_{\beta \gamma}$
can be neglected assuming high order gradients are negligibly small.

A detailed comparison of the terms in this equation to the target
momentum equation \e(\ref{eq_qs_momentum}) gives the identifications
made in \es(\ref{eq_f_anals}). Note that the isotropic terms may be
incorporated into the scalar pressure~\cite{deGennes:1993}.
Comparison of the remaining stress tensor terms  allows the forcing
term \e(\ref{eq_lb_mom_foreterm}) to be defined. This completes the
momentum analysis.

\subsection{Order Tensor Evolution} \label{sec:derive:ce:ord} The moments
of the order distribution function are defined to be
\begin{equation}\label{eq:ch4:31}
    \left.
    \begin{array}{lllccc}
        \sum_{i}g_{i\alpha \beta}^{(0)}
            \left[
            \begin{array}{ccccc}
                1 \\
                c_{i\gamma}
            \end{array}
            \right] &=& \left[
            \begin{array}{ccccc}
                 S_{\alpha \beta} \\
                  u_{\gamma}S_{\alpha \beta}
            \end{array}
            \right]
            \\
            \sum_{i}g_{i\alpha \beta}^{(n)}
            \left[
            \begin{array}{ccccc}
                1 \\
                c_{i\gamma}
            \end{array}
            \right] &=& \left[
            \begin{array}{ccccc}
                0 \\
                \Omega^{(n)}_{\alpha \beta \gamma}
            \end{array}
            \right] \;\;,\;\;\;\; n > 0
    \end{array}
    \right\}
\end{equation}
It should be noted that we adopt a unit trace order tensor in these
moment definitions in order to be consistent with
~\cite{Care:2003.061703}. However, since the momentum and order are
now in separate algorithms, we could equally well have defined zero
trace moment definitions as in~\cite{Denniston:2000.481}.

Using a Taylor expansion on the left hand side of lattice evolution
equation (Eq.(\ref{eq_g_lbgk})) we obtain:
\begin{equation}\label{eq:ch4:32}
    \begin{array}{ccc}
        \triangle t \partial_{t}g_{i\mu \nu}+\frac{\triangle t^{2}}{2}\partial_{t}\partial_{t}g_{i\mu \nu} +
        \triangle t c_{i{\alpha}}\partial_{\alpha}g_{i\mu \nu} \\~~~+ \triangle t^{2}c_{i\alpha}\partial_{t}\partial_{\alpha}g_{i\mu \nu}
        +\frac{\triangle t^{2}}{2}c_{i\alpha }c_{i\beta}\partial_{\alpha} \partial_{\beta}g_{i\mu \nu}
        \\= -\frac{1}{\tau_{_{Q}}}\left( g_{i\mu \nu}-g_{i\mu \nu}^{(eq)} \right) + \chi_{i\mu \nu}
    \end{array}
\end{equation}
We suppose the as yet unknown forcing term $\chi_{i\mu \nu}$ will be
 dependent on the gradient in both $\mathbi{u}$ and $\mathbi{Q}$ and can be expanded as $\chi_{i\mu
\nu}=\varepsilon \chi_{i\mu \nu}^{(1)}+\varepsilon^{2}\chi_{i\mu
\nu}^{(2)}$.  We augment \es(\ref{eq:ch4:22}) with the expansion
\begin{equation}\label{eq:ch4:33}
    \begin{array}{ccc}
        g_{i}=g_{i\mu \nu}^{(0)}+\varepsilon g_{i\mu \nu}^{(1)}+\varepsilon^{2}g_{i\mu \nu}^{(2)}
    \end{array}
\end{equation}
Substituting into the Taylor expansion \e(\ref{eq:ch4:32}), we find
$O(\varepsilon ^{0})$
\begin{eqnarray}
    \label{eq:ch4:35}
     & g_{i\mu \nu}^{(0)} = g_{i\mu \nu}^{(eq)}
\end{eqnarray}
$O(\varepsilon ^{1}) $
\begin{eqnarray}
    \label{eq:ch4:36}
    & -\tau_{_{Q}}\triangle t (\partial_{t_{1}}+
    c_{i\alpha_{1}}\partial_{\alpha_{1}} )g_{i\mu \nu}^{(0)} + \tau_{_{Q}}\chi_{i\mu \nu}^{(1)} = g_{i\mu \nu}^{(1)}
\end{eqnarray}
$O(\varepsilon ^{2}) $
\begin{eqnarray}
    \label{eq:ch4:37}
    \left(\frac{1}{2}-\tau_{_{Q}}\right) \triangle t\left( \partial_{t_{1}}+c_{i\alpha}\partial_{\alpha_{1}} \right)g_{i\mu \nu}^{(1)} \nonumber \\- \tau_{_{Q}}\triangle t \left( \partial_{t_{2}}+c_{i\alpha}\partial_{\alpha_{2}} \right) g_{i\mu \nu}^{(0)}
     \nonumber \\ -\frac{\tau_{_{Q}}\triangle t }{2}\chi_{i\mu \nu}^{(1)} +\tau_{_{Q}}\chi_{i\mu \nu}^{(2)} & = & g_{i\mu \nu}^{(2)}
\end{eqnarray}
in which we have used the $O(\varepsilon^{1})$ result of
\e(\ref{eq:ch4:36}) to replace a term of the form
$(\partial_{t_{1}}+c_{i\alpha}\partial_{\alpha_{1}})g_{i\mu
\nu}^{(0)}$ in the $O(\varepsilon^{2})$ result.

Taking the zeroth moment  of the first order \e(\ref{eq:ch4:36})
expansion and using \e(\ref{eq:ch4:31}) gives
\begin{equation}\label{eq:ch4:38}
    \partial_{t_{1}}\left(  S_{\mu \nu } \right)+\partial_{\alpha_{1}}\left(  u_{\alpha}S_{\mu \nu}\right)=\sum_{i}\frac{\chi_{i\mu \nu}^{(1)}}{\triangle t}
\end{equation}
which may be written in terms of $\mathbi{Q}$ as
\begin{equation}\label{eq:ch4:39}
     \partial_{t_{1}}\left( Q_{\mu \nu } \right)+ \partial_{\alpha_{1}}\left( u_{\alpha}Q_{\mu \nu}\right)=\frac{3}{2\triangle t} \sum_{i}\chi_{i\mu \nu}^{(1)}
\end{equation}
Similarly, the zeroth moment  of the second order expansion gives
\begin{eqnarray}\label{eq:ch4:40}
   \left( 1-\frac{1}{2\tau_{_{Q}}} \right) \partial_{\alpha_{1}} \Omega_{\alpha \mu \nu}^{(1)} + \frac{2 }{3} \partial_{t_{2}}Q_{\mu \nu }+\frac{2}{3} \partial_{\alpha_{2}}\left( u_{\alpha}Q_{\mu \nu}\right)&=& \nonumber \\
    +\frac{1}{\triangle t}\sum_{i}\chi_{i\mu \nu }^{(2)}-\frac{1}{2} \sum_{i}\chi_{i\mu
    \nu}^{(1)}~~~~~
\end{eqnarray}

We now need to evaluate $\Omega_{\alpha \mu \nu}^{(1)}$ by obtaining
an expression for $g_{i\mu \nu}^{(1)}$ in \e(\ref{eq:ch4:36}). We
use Eq.~(\ref{eq_g_equilib}) to $O(\underline{\mathbi{u}})$ in
Eq.(\ref{eq:ch4:36}). Taking the first moment of this gives
\begin{eqnarray}\label{eq:ch4:41}
  \Omega_{\gamma \mu \nu}^{(1)}= \tau_{_{Q}}\sum_{i}c_{i\gamma}\chi_{i\mu \nu }^{(1)}
    -\tau_{_{Q}}\triangle t\left[  S_{\mu \nu}\partial_{t_{1}}(u_{\gamma})\right.\nonumber \\
    \left.+u_{\gamma}\partial_{t_{1}}( S_{\mu \nu}) + \partial_{\gamma_{1}}( S_{\mu \nu}c_{s}^{2})
    \right]
\end{eqnarray}
Using the result \e(\ref{eq:ch4:38}), we may replace
$\partial_{t_{1}}( S_{\mu \nu})$ and find
\begin{eqnarray}\label{eq:ch4:42}
        \Omega_{\gamma \mu \nu}^{(1)}=-\tau_{_{Q}}\triangle t \biggl[  S_{\mu \nu}\partial_{t_{1}}(u_{\gamma})-u_{\gamma}\partial_{\alpha_{1}}( u_{\alpha}S_{\mu \nu}) \biggr. \nonumber \\
        \left. + \frac{u_{\gamma}}{\triangle t}\sum_{i}\chi_{i\mu \nu}^{(1)} + \partial_{\gamma_{1}}( S_{\mu \nu}c_{s}^{2})
        \right]\nonumber \\
        + \tau_{_{Q}}\sum_{i}c_{i\gamma}\chi_{i\mu \nu }^{(1)}~~~
\end{eqnarray}
We may use the result obtained in text above Eq.~(\ref{eq:ch4:27})
to find:
\begin{equation}\label{eq:ch4:43}
    \partial_{t_{1}}u_{\beta}=-\frac{c_{s}^{2}\partial_{\beta_{1}}\rho-\tau_{_{P}}\partial_{\gamma_{1}}F_{\beta \gamma}c_{s}^{2}+u_{\beta}\partial_{\gamma_{1}}(\rho u_{\gamma})}{\rho}
\end{equation}
and hence from \e(\ref{eq:ch4:42}) and the incompressibility
condition we find
\begin{eqnarray}\label{eq:ch4:44}
        \Omega_{\gamma \mu \nu}^{(1)}=-\tau_{_{Q}}\triangle t\left( -\frac{ S_{\mu \nu}\partial_{\beta_{1}}(F_{\beta \gamma})c_{s}^{2}\tau_{_{P}}}{\rho }-u_{\gamma}u_{\alpha}\partial_{\alpha_{1}}( S_{\mu \nu}) \right.\nonumber \\
        \left.+ \frac{u_{\gamma}}{\triangle t}\sum_{i}\chi_{i\mu \nu}^{(1)} + c_{s}^{2}\partial_{\gamma_{1}}( S_{\mu \nu})  \right) + \tau_{_{Q}}\sum_{i}c_{i\gamma}\chi_{i\mu \nu
        }^{(1)}~~~~~
\end{eqnarray}

Upon converting $\mathbi{S}$ to $\mathbi{Q}$ Eq.~(\ref{eq:ch4:44})
is inserted in the earlier second order zeroth moment
Eq.~(\ref{eq:ch4:40}) giving:
\begin{widetext}

\begin{eqnarray}\label{eq:ch4:45}
             \partial_{t_{2}}Q_{\mu \nu}+ \partial_{\alpha_{2}}(u_{\alpha} Q_{\mu \nu})+\frac{3}{2}\left( 1-\frac{1}{2\tau_{_{Q}}} \right)
        \left[ \frac{2 \tau_{_{P}}\tau_{_{Q}}c_{s}^{2} \triangle t\partial_{\alpha_{1}}(Q_{\mu \nu}\partial_{\beta_{1}}(F_{\beta \alpha})}{3\rho }  \right. \nonumber\\
        \left. +\frac{ \tau_{_{P}}\tau_{_{Q}}\triangle t \delta_{\mu \nu}c_{s}^{2}}{3\rho} \partial_{\alpha_{1}} (\partial_{\beta_{1}} F_{\beta \alpha} ) +
        \frac{2\tau_{_{Q}}\triangle t}{3}\partial_{\alpha _{1}}(u_{\alpha}u_{\gamma}\partial_{\gamma_{1}}Q_{\mu \nu}) - \tau_{_{Q}}\partial_{\alpha_{1}}(u_{\alpha}\sum_{i}\chi_{i\mu \nu}^{(1)}) \right.\\
        \left. - \frac{2\tau_{_{Q}}\triangle t c_{s}^{2}}{3}\partial_{\alpha _{1}}(\partial_{\alpha _{1}}Q_{\mu \nu}) + \tau_{_{Q}}\partial_{\alpha_{1}}(\sum_{i}c_{i\alpha}\chi_{i\mu \nu}^{(1)}) \right] = -\frac{3}{4}\sum_{i}\chi_{i\mu \nu}^{(1)}+\frac{3}{2\triangle t}\sum_{i}\chi_{i\mu
        \nu}^{(2)}\nonumber
\end{eqnarray}

\end{widetext}

In order to simplify Eq. \ref{eq:ch4:45} we omit terms which include
third order gradients in either $\mathbi{u}$  or $\mathbi{Q}$.
Further in the limit $M=\frac{|\underline{u}|}{c_{s}^{1}} \ll 1 $,
which holds for low $Re$ LCs, we may omit the term which includes
the product $\mathbi{u u}$.

Recombining the $O(\varepsilon ^{1})$ Eq.~(\ref{eq:ch4:39}) and
$O(\varepsilon ^{2})$ Eq.~(\ref{eq:ch4:45}) expansion we obtain:
\begin{eqnarray}\label{eq:ch4:46}
    \partial_{t}Q_{\mu \nu} + u_{\alpha }\partial_{\alpha}Q_{\mu \nu}=\frac{c_{s}^{2}}{2}\left( 2\tau_{_{Q}} -1 \right) \triangle t \partial_{\alpha}\left( \partial_{\alpha}Q_{\mu \nu} \right)
        \nonumber \\ -\frac{3 }{4}\sum_{i}\chi_{i\mu \nu}^{(1)} + \frac{3}{2\triangle t  }\sum_{i}\chi_{i\mu
        \nu}^{(2)}~~~~
\end{eqnarray}

A  comparison of the terms in this equation and the target order
equation (Eq.~(\ref{eq_new_order_evolu})) gives the identification
made  in Eq.~(\ref{eq_g_anals}). We now compare the remaining terms
with the force terms in Eq.~(\ref{eq:ch4:46}). We make the
assumption that the forcing term must be introduced  at
$O(\varepsilon^{2})$ since it is gradient dependent; we therefore
choose $\sum_{i}\chi_{i\mu \nu}^{(1)}=0$.  This allows us to make
the identification for the forcing term, $\chi_{i\alpha\beta}$,
given in Eq.~(\ref{eq_chi_force}).

\section{\label{parameters}Algorithm parameters}
Here we give details on the relations between EL material
coefficients and the $\mathbi{Q}$ tensor coefficients. We use
standard EL notations as given from~\cite{deGennes:1993}. The
details are obtained by using Eq.~(\ref{eq_q_tensor}) in
Eqs.~(\ref{eq_qs_momentum}) and~(\ref{eq_qs_order}). Note $S_{0}$
stands for the equilibrium order parameter, not simulation evolved
order parameter, $S$.
\begin{widetext}
\begin{equation}\label{eq_appen_1}
    \begin{array}{c}
        \beta _{1}=\frac{4\alpha _{1}}{9S_{0}^{2}} \;,\;\;\;\;\; \beta _{4}=\alpha _{4}+\frac{\alpha _{5}+\alpha _{6}}{3}
        \;,\;\;\;\;\; \beta _{5}=\frac{2\alpha _{5}}{3S_{0}} \;,\;\;\;\;\; \beta _{6}=\frac{2\alpha _{6}}{3S_{0}} \\
         \mu _{1}=\frac{2(\alpha _{3}-\alpha _{2})}{9S_{0}^{2}}=\frac{2\gamma_{1}}{9S_{0}^{2}} \;,\;\;\;\;\; \mu _{2}=\frac{2(\alpha _{2}+\alpha _{3})}{3S_{0}}=\frac{2\gamma_{2}}{3S_{0}}=\beta _{6}-\beta
         _{5} \;,\;\;\;\;\; \mu _{s}=\frac{2\gamma _{s}}{9S_{0}^{2}} \\
        L_{1}=\frac{2}{27S_{0}^{2}}\left( 3K_{22}+K_{33}-K_{11}\right) \;,\;\;\;\;\;
        L_{2}=\frac{4}{9S_{0}^{2}}\left( K_{11}-K_{22}-K_{24}\right) \;,\;\;\;\;\;
        L_{3}=\frac{4}{9S_{0}^{2}} K_{24} \;,\;\;\;\;\;
        L_{4}=\frac{4}{27S_{0}^{3}}\left( K_{33}-K_{11}\right) \\
        \alpha _{F}=\frac{4}{3}a(T-T^{\ast }) \;,\;\;\;\;\;  \beta _{F}=\frac{4}{3}B \;,\;\;\;\;\; \gamma _{F}=\frac{4}{9}C\;,\;\;\;\;\;
        S_{IN}=\frac{B}{2C} \;,\;\;\;\;\; T^{\ast }=T_{IN}-\frac{B^{2}}{4aC} \\
        \xi_{1}=\frac{2}{9S_{0}}(e_{11}+2e_{33}) \;,\;\;\;\;\;
        \xi_{2}=\frac{4}{9S_{0}^{2}}(e_{11}-e_{33}) \\
    \end{array}
\end{equation}

Then the relation of the $\mathbi{Q}$ tensor coefficients to both
momentum algorithm and the order algorithms are

\begin{equation}\label{eq_appen_2}
    \begin{array}{cc}
        \mu_{1}^{\prime _{P}}=\frac{ \mu_{1} \rho^{\prime} c_{s}^{2}(2\tau_{_{P}}-1)\triangle t^{\prime }}{2\eta_{eff}}&
        \mu_{1}^{\prime _{Q}}=\mu_{1}^{\prime _{P}}\left (\frac{Er}{Re} \right)\\
        \mu_{2}^{\prime _{P}}=\frac{\mu_{2}\mu_{1}^{\prime _{P}}}{\mu_{1}}&\mu_{2}^{\prime _{Q}}=\mu_{2}^{\prime _{P}}\left (\frac{Er}{Re} \right)\\
        \beta_{1}^{\prime _{P}}=\frac{\beta_{1}\mu_{1}^{\prime _{P}}}{\mu_{1}}&\beta_{1}^{\prime _{Q}}=\beta_{1}^{\prime _{P}}\left (\frac{Er}{Re} \right)\\
        \beta_{4}^{\prime _{P}}=\frac{\beta_{4}\mu_{1}^{\prime _{P}}}{\mu_{1}}&\beta_{4}^{\prime _{Q}}=\beta_{4}^{\prime _{P}}\left (\frac{Er}{Re} \right)\\
        \beta_{5}^{\prime _{P}}=\frac{\beta_{5}\mu_{1}^{\prime _{P}}}{\mu_{1}}&\beta_{5}^{\prime _{Q}}=\beta_{5}^{\prime _{P}}\left (\frac{Er}{Re} \right)\\
        \beta_{6}^{\prime _{P}}=\frac{\beta_{6}\mu_{1}^{\prime _{P}}}{\mu_{1}}&\beta_{6}^{\prime _{Q}}=\beta_{6}^{\prime _{P}}\left (\frac{Er}{Re} \right)\\
    \end{array}
\end{equation}

\begin{equation}\label{eq_appen_3}
    \begin{array}{cc}
        L_{1}^{\prime _{P}}=L_{1}^{\prime _{Q}}\left( \frac{Er}{Re}  \right) ^{-2}&L_{1}^{\prime _{Q}}=\mu_{1}^{\prime _{Q}}\frac{c_{s}^{2}}{2}(2\tau_{_{Q}}-1)\triangle t^{\prime } \\
        L_{2}^{\prime _{P}}=L_{2}^{\prime _{Q}}\left( \frac{Er}{Re}  \right) ^{-2}&L_{2}^{\prime _{Q}}=L_{2}\frac{L_{1}^{\prime _{Q}}}{L_{1}}\\
        L_{3}^{\prime _{P}}=L_{3}^{\prime _{Q}}\left( \frac{Er}{Re}  \right) ^{-2}&L_{3}^{\prime _{Q}}=L_{3}\frac{L_{1}^{\prime _{Q}}}{L_{1}}\\
        L_{4}^{\prime _{P}}=L_{4}^{\prime _{Q}}\left( \frac{Er}{Re}  \right) ^{-2}&L_{4}^{\prime _{Q}}=L_{4}\frac{L_{1}^{\prime _{Q}}}{L_{1}}\\
        \alpha_{F}^{\prime _{P}}=\alpha_{F}^{\prime _{Q}}\left( \frac{Er}{Re}  \right) ^{-2}&\alpha_{F}^{\prime _{Q}}=\alpha_{F}\frac{\bar{L}^{2}L_{1}^{\prime _{Q}}}{\bar{L}^{\prime 2}L_{1}}\\
        \beta_{F}^{\prime _{P}}=\beta_{F}^{\prime _{Q}}\left( \frac{Er}{Re}  \right) ^{-2}&\beta_{F}^{\prime _{Q}}=\beta_{F}\frac{\alpha_{F}^{\prime _{Q}}}{\alpha_{F}}\\
        \gamma_{F}^{\prime _{P}}=\gamma_{F}^{\prime _{Q}}\left( \frac{Er}{Re}  \right) ^{-2}&\gamma_{F}^{\prime _{Q}}=\gamma_{F}\frac{\alpha_{F}^{\prime _{Q}}}{\alpha_{F}}\\
    \end{array}
\end{equation}

\begin{equation}\label{eq_appen_4}
    \begin{array}{cc}

        \triangle t_{P}=\triangle t_{Q} \left( \frac{Er}{Re}  \right) ^{-1}&\triangle t_{Q}(=\triangle t_{n}=\triangle t_{s})=\frac{c_{s}^{2}}{2}(2\tau_{_{Q}}-1)\frac{\triangle t^{\prime 2}\gamma_{1}\bar{L}^{2}}{K_{22}\bar{L}^{\prime ^{2}}}\\
        E^{\prime _{P}}=E^{\prime _{Q}} \left( \frac{Er}{Re}  \right) ^{-3}&E^{\prime _{Q}}=\sqrt{\frac{\epsilon_{0}\triangle \epsilon_{a}
        E^{2}9S_{0}^{2}\mu_{1}^{\prime _{Q}}\triangle t_{Q}}{2\gamma_{1}\triangle t^{\prime }\epsilon_{0}^{\prime _{Q}}\triangle \epsilon_{a}^{\prime _{Q}}
        }}\\
        \mu_{s}^{\prime _{P}}=\mu_{s}^{\prime _{Q}} \left( \frac{Er}{Re}  \right) ^{-1}&\mu_{S}^{\prime _{Q}}=\frac{L_{1}^{\prime _{Q}}\gamma_{S}\triangle^{\prime } t\bar{L}}{K_{22}\triangle t_{Q}\bar{L}^{\prime }}\\
        W^{\prime _{P}} = W^{\prime _{Q}} \left( \frac{Er}{Re}  \right) ^{-2}&W^{\prime _{Q}}=\frac{2W\mu_{S}^{\prime _{Q}}\triangle t_{Q}}{\gamma_{S}\triangle t^{\prime }}\\
        e_{11}^{\prime _{P}}=e_{11}^{\prime _{Q}} \left( \frac{Er}{Re}  \right) ^{-1}&e_{11}^{\prime _{Q}}=\frac{e_{11}9S_{0}^{2}\mu_{S}^{\prime _{Q}}\triangle t_{Q}E}{2\gamma_{S}\triangle t^{\prime }E^{\prime _{Q}}}\\
        e_{33}^{\prime _{P}}=e_{33}^{\prime _{Q}} \left( \frac{Er}{Re}  \right) ^{-1}&e_{33}^{\prime _{Q}}=\frac{e_{33}9S_{0}^{2}\mu_{S}^{\prime _{Q}}\triangle t_{Q}E}{2\gamma_{S}\triangle t^{\prime }E^{\prime _{Q}}}\\
    \end{array}
\end{equation}
\end{widetext}
Note that in the definition of $\triangle t_Q$ we have used an
elastic constant which is characteristic of a simple twisted nematic
cell to illustrate the mapping of variables, plus $\epsilon
_{0}^{\prime _{Q}}=1$ and $\triangle t^{\prime^{Q}}=\triangle
t^{\prime^{P}}=\triangle t^{\prime}$ and
$L^{\prime^{Q}}=L^{\prime^{P}}=L^{\prime}$ and
$\mathbi{v}=\mathbi{v}^{\prime_{P}}\frac{2\bar{L}^{\prime_{P}}\eta_{eff}}{\rho
\bar{L}c_{s}^{2}\left( 2\tau_{_{P}}-1\right)}$ and
$\phi=\phi^{\prime_{Q}}\sqrt{\frac{2\gamma_{1}\bar{L}^{2}}
{9S_{0}^{2}\mu_{1}^{\prime_{Q}}\epsilon_{0}\triangle
t_{Q}\bar{L}^{\prime2}} }$.

The material parameters used for simulations in
\S\ref{device_results} are: $(K_{11}=10,~K_{22}=7,~ K_{33}=14,~
K_{24}=5) \times10^{-12} \rm{kg~m~s}^{-2}$,~$(\alpha
_{1}=-11,~\alpha _{2}=-102,~\alpha _{3}=-5,~\alpha _{4}=74,~\alpha
_{5}=84,~\alpha _{6}=-23)
\times10^{-3}\rm{kg}~^{-1}~\rm{s}^{-1}$,~$\rho =
1.01\times10^{3}\rm{kg~m}^{-3}$,~$T=T_{IN}-4(T_{IN}-T^{\star})\;\rm{K}$,~$a=65000\:\rm{J\:m^{-3}\:
K^{-1}}$,~ $B=530000\:\rm{J\:m^{-3}}$,~$C=980000\:\rm{J\:m^{-3}}$
,~$W=7\times10^{-4}kg~s^{-2}$,~ $\epsilon _{a} = 10.3$,~$\epsilon
_{\gamma \gamma
}=18$,~$\frac{1}{2}(e_{11}+e_{33})=2\times10^{-11}A~S~m^{-1}$,~
$l_{S}=10^{-7}\rm{m}$. Other specific constants are provided in the
figure captions. The parameters used represent a hybrid of commonly
used materials; they do not correspond to a specific material since
a complete set of material parameters does not exist in the
literature for one material.
\newpage


\bibliographystyle{apsrev}

\end{document}